\pdfoutput=1

\documentclass[11pt,twoside,a4paper,cmspaper,final,collab]{cms-tdr}
\def\svnVersion{651597e}\def\svnDate{2020/12/10}\def\cmsCernNoTag{CERN-EP-2020-040}\def\cmsCernDate{\today}\def\cmsMessage{"Published in the Journal of High Energy Physics as \href{http://dx.doi.org/10.1007/JHEP12(2020)082}{\doi{10.1007/JHEP12(2020)082}.}"}

\begin{document}\cmsNoteHeader{SMP-19-003}

\hyphenation{had-ron-i-za-tion}
\hyphenation{cal-or-i-me-ter}
\hyphenation{de-vices}
\newcommand{\NLOJET} {{\textsc{nlojet++}}\xspace}
\newcommand{\FASTNLO} {{\textsc{fastnlo}}\xspace}
\newcommand{\HERWIGS} {{\textsc{herwig7}}\xspace}
\newcommand{\pQCD}{{p\text{QCD}}\xspace}

\cmsNoteHeader{SMP-19-003} 

\title{Dependence  of inclusive jet production on the anti-\kt distance parameter in pp collisions at $\sqrt{s} = 13$\TeV}

\author*[inst1]{Suman Chatterjee}

\date{\today}

\abstract{
  The dependence of inclusive jet production
  in proton-proton collisions with a center-of-mass energy of 13\TeV
  on the distance parameter $R$ of the anti-\kt algorithm
  is studied using data corresponding to integrated luminosities up to 35.9\fbinv 
  collected by the CMS experiment in 2016.
  The ratios of the inclusive cross sections as functions of 
  transverse momentum \pt and rapidity $y$, 
  for $R$ in the range 0.1 to 1.2
  to those using $R=0.4$ 
  are presented in the region $84< \pt <1588\GeV$ and $\abs{y}<2.0$.  The results are compared to calculations at leading and next-to-leading order in the strong coupling constant using different parton shower models. 
  The variation of the ratio of cross sections with $R$ is well described by calculations including a parton shower model, but not by a leading-order quantum chromodynamics calculation including nonperturbative effects.
  The agreement between the data and the theoretical predictions for the ratios of cross sections is significantly improved
  when next-to-leading order calculations with nonperturbative effects are used. 
}

\hypersetup{
pdfauthor={CMS Collaboration},%
pdftitle={Dependence  of inclusive jet production on the anti-kt distance parameter in pp collisions at sqrt(s) = 13 TeV},%
pdfsubject={CMS},%
pdfkeywords={CMS, physics, QCD, inclusive jets}}

\maketitle

\section{Introduction}

Quantum chromodynamics (QCD) is a gauge theory describing the strong interaction
between partons (quarks and gluons).
Jets are reconstructed using hadron particles produced by the fragmentation of partons in collisions~\cite{PhysRevLett.39.1436}.
Thus jets approximate the original partons created in short-distance scatterings.
The production cross sections for high transverse momentum (\pt) partons can be calculated
using perturbative QCD (\pQCD).
Specifically, predictions for hadron production in proton-proton collisions require models for
parton showering~\cite{Sjostrand:1985xi,Gottschalk:1986bk,Marchesini:1983bm}
and nonperturbative (NP) effects such as
hadronization~\cite{Dokshitzer:1995zt} and underlying event (UE)~\cite{PhysRevD65092002}.
When the fixed-order prediction in \pQCD is not adequate, higher-order terms must be included using resummation methods~\cite{Kidonakis:1998bk,SEYMOUR1998269,Liu:2017pbb}.

The results of measurements of inclusive jet production cross sections 
for proton-proton collisions are typically presented using the anti-\kt jet algorithm~\cite{anti-kT}
characterized by a distance parameter $R$, which is a measure of the jet size in rapidity-azimuth plane.
Anti-\kt jets with distance parameter $R$ are referred to as AK$n$ jets, where $R = 0.1 n$.
The CMS Collaboration~\cite{CMS_ex} has reported measurements 
at center-of-mass energy ($\sqrt{s}$ ) of 7\TeV~\cite{sigma7} and 8\TeV~\cite{sigma8} 
using AK5 and AK7 jets.
The CMS results at $\sqrt{s} = 13$\TeV 
for AK4 and AK7 jets
are reported in Ref.~\cite{sigma13}.
After the application of a correction for NP and electroweak effects,
the results for AK7 jets are well described by next-to-leading order (NLO)
calculations based on the {\NLOJET}~\cite{Nagy:2003tz} program
used in the \FASTNLO software package~\cite{FNLO_ref}.
The prediction from the \POWHEG~\cite{Powheg_ref} generator, which also computes matrix elements at NLO
and 
is used with parton showering
simulated with {\PYTHIA}8~\cite{pythia} or {\HERWIGpp}~\cite{herwig},
describes results well for both AK7 and AK4 jets.
However, the ATLAS Collaboration has measured the production cross sections for both AK4 and AK6 jets and finds a discrepancy between the measured results and the \POWHEG prediction~\cite{ATLAS_8TeV}.
The ATLAS Collaboration has also compared the measurements of inclusive jet production at $\sqrt{s} = 13$\TeV  with the next-to-next-to-leading order (NNLO) prediction in {\pQCD}~\cite{ATLAS_13TeV}.

The measurement of a jet production cross section as a function
of the distance parameter is sensitive to the details of the theoretical modeling
of the perturbative and NP processes involved in the evolution of the partons.
The measurement of the ratio of cross sections with two jet sizes was first performed by the ALICE Collaboration with AK2 and AK4 jets~\cite{Abelev:2013fn}.
A similar study was also produced by the CMS Collaboration with AK5 and AK7 jets~\cite{Mikko}.
We explore this topic further in the present paper by extending the measurement to various values of jet size.
Recently, ALICE Collaboration has also measured both the absolute cross sections of inclusive jet production and the ratio of cross sections for $R = 0.1$--0.6 in $20 < \pt < 140\GeV$~\cite{Acharya:2019jyg}. 
Dependence of inclusive jet production on the distance parameter is also studied in detail in Ref.~\cite{Bellm:2019yyh}.

Quarks and gluons radiate secondary gluons that can be emitted outside of the catchment area of the jet definition, which is the region in rapidity-azimuth plane contributing to the jet. 
This lost \pt
is calculated using a QCD splitting function, 
with the leading-order (LO) result~\cite{Aversa:1988vb,deFlorian:2007fv,fragmentation}
in the small-$R$ approximation ($R \ll 1$) 
\begin{equation}
(\delta \pt)_{\mathrm{q}} = -C_{\mathrm{F}}\frac{\alpS \pt}{\pi} \ln \biggl(\frac{1}{R}\biggr)\biggl(2 \ln 2 - \frac{3}{8}\biggr) + \mathcal{O}(\alpS) ,
\end{equation}
for quark-initiated jets and
\begin{equation}
(\delta \pt)_{\mathrm{g}} = - \frac{\alpS \pt}{\pi} \ln \biggl(\frac{1}{R}\biggr)\left[C_{\mathrm{A}}(2 \ln 2-\frac{43}{96})+T_{\mathrm{R}}n_{\mathrm{f}}\frac{7}{48}\right] + \mathcal{O}(\alpS)  , 
\end{equation}
for gluon-initiated jets.
 
Here $C_{\mathrm{F}}(=\frac{4}{3})$ and $C_{\mathrm{A}}(=3)$ are the Casimir factors for quarks and
gluons respectively, $T_{\mathrm{R}}(=\frac{1}{2})$ is the SU(3) quantum number, and
$n_{\mathrm{f}}$ is the number of active quark flavors.
Larger values of $R$ capture a larger fraction of the radiation.

Properties of jets are also modified by hadronization, an NP process describing the transition of partons into hadrons.
As described in Ref.~\cite{hadronisation}, some theoretical models parameterize the effect of hadronization by taking $\alpS (\mu) = \mu_{\mathrm{l}}\delta(\mu-\mu_{\mathrm{l}})$, where $\mu_{\mathrm{l}}$ is commensurate with the  Landau pole, yielding
\begin{equation}
(\delta \pt)_{\mathrm{had}} \simeq -\frac{2 C {\mathcal{A}}(\mu_{\mathrm{l}})}{\pi R} + \mathcal{O}(R) ,
\end{equation}
in the small-$R$ limit, where $C = C_{\mathrm{F}} (C_{\mathrm{A}}) $  
for quark (gluon) initiated jets, 
and ${\mathcal{A}}(\mu_{\mathrm{l}})$ is related to the scale appearing in the calculations of hadronization. 
Losses are again minimized at larger values of $R$.

The algorithm defining the jets
can also select particles  from the underlying event, 
which in general involves low momentum transfer.
These particles typically
have low \pt. The energy density ($\Lambda_{\mathrm{UE}}$ per unit $y$) from these sources 
is approximately uniform over the jet area, and
their contribution to the jet \pt is approximately given~\cite{jet-area, fragmentation} by
\begin{equation}
(\delta \pt)_{\mathrm{UE}} \simeq \frac{1}{2} \Lambda_{\mathrm{UE}}R^{2}  , 
\end{equation}
for small $R$ values.

Since, as discussed above, the
contributions of various perturbative and NP effects depend on the jet
size, and because radiation and hadronization are different for jets
initiated by quarks and by gluons, comparisons of jets with different
cone sizes yield information about these processes, and can be used to
improve theoretical calculations.

In this paper, we present measurements of the ratio of the
cross section for inclusive
anti-\kt jets with distance parameters of $R=0.1,0.2,\ldots,1.2$ 
to that of AK4 jets.
The results are compared with
predictions from different Monte Carlo (MC) generators,
involving matrix element calculations at different orders and utilizing
different parton shower and hadronization models.
Predictions for cross section ratios have also been obtained
using a \pQCD calculation at NLO that uses the following convention
\begin{equation}
\text{Ratio}(R, \pt) = \frac{\left(\frac{\rd\sigma^{(R)}}{\rd\pt} - \frac{\rd\sigma^{(0.4)}}{\rd\pt}\right)} {\frac{\rd\sigma^{(0.4)}}{\rd\pt}} + 1  , 
\label{rad_diff}
\end{equation}
where $R$ is the anti-\kt jet distance parameter,
and $R=0.4$ is taken as the reference jet size. 
The terms in Eq.~\eqref{rad_diff} are differential cross sections for three-jet production
and are calculated at fixed-order
using \NLOJET with terms up to $\alpha_{S}^4$~\cite{Nagy:2001fj,Soyez:2011np}.
Measurements are restricted to $\pt < 1588$\GeV because of the large experimental uncertainty 
in the calibration of high energy jets, which was not optimized for the cross section ratios.

\section{The CMS detector} \label{sec_CMS}

The central feature of the CMS apparatus is a superconducting solenoid of 6\unit{m} internal diameter, providing a magnetic field of 3.8\unit{T}. Within the solenoid volume are a silicon pixel and strip tracker, a lead tungstate crystal electromagnetic calorimeter (ECAL), and a brass and scintillator hadron calorimeter (HCAL), each composed of a barrel and two endcap sections. Forward calorimeters extend the pseudorapidity ($\eta$) coverage provided by the barrel and endcap detectors. Muons are detected in gas-ionization chambers embedded in the steel flux-return yoke outside the solenoid. A more detailed description of the CMS detector, together with a definition of the coordinate system used and the relevant kinematic variables, can be found in Ref.~\cite{CMS_ex}.

The silicon tracker measures charged particles within the range $\abs{\eta} < 2.5$. It consists of 1440 silicon pixel and 15\,148 silicon strip detector modules. For nonisolated particles of $1 < \pt < 10\GeV$ and $\abs{\eta} < 1.4$, the track resolutions are typically 1.5\% in \pt and 25--90 (45--150)\mum in the transverse (longitudinal) impact parameter \cite{TRK-11-001}.
In the barrel section of the ECAL, an energy resolution of about 1\% is achieved for unconverted or late-converting photons that have energies in the range of tens of GeV. The remaining barrel photons have a resolution of about 1.3\% up to $\abs{\eta} = 1$, rising to about 2.5\% at $\abs{\eta} = 1.4$. In the endcaps, the resolution of unconverted or late-converting photons is about 2.5\%, while the remaining endcap photons have a resolution between 3 and 4\%~\cite{CMS:EGM-14-001}. 
In the region $\abs{\eta} < 1.74$, the HCAL cells have widths of 0.087 in $\eta$ and 0.087 radians in azimuth ($\phi$). In the $\eta$-$\phi$ plane, and for $\abs{\eta} < 1.48$, the HCAL cells map on to $5 \times 5$ arrays of ECAL crystals to form calorimeter towers projecting radially outwards from close to the nominal interaction point. For $\abs{\eta} > 1.74,$ the coverage of the towers increases progressively to a maximum of 0.174 in $\Delta \eta$ and $\Delta \phi$~\cite{HCALTDR1997}. Within each tower, the energy deposits in ECAL and HCAL cells are summed to define the calorimeter tower energies, subsequently used to provide the energies and directions of hadronic jets.
When combining information from the entire detector, the jet energy resolution amounts typically to 15\% at 30\GeV, 10\% at 100\GeV, and 5\% at 1\TeV at $\abs{\eta}<0.5$, while at $\abs{\eta}=2.0$ the jet energy resolution increases by 1--2\% at low {\pt}~\cite{CMS-DP-2016-020}.

Events of interest are selected using a two-tiered trigger system~\cite{Khachatryan:2016bia}. The first level, composed of custom hardware processors, uses information from the calorimeters and muon detectors to select events at a rate of around 100\unit{kHz}. The second level, known as the high-level trigger (HLT), consists of a farm of processors running a version of the full event reconstruction software optimized for fast processing that reduces the event rate to around 1\unit{kHz} before data storage.

\section{Jet reconstruction} \label{sec_jetreco}

The CMS particle-flow algorithm~\cite{PFA} reconstructs and identifies each individual particle in an event with an optimized combination of information from the various elements of the CMS detector. The energy of photons is obtained from the ECAL measurement. The energy of electrons is determined from a combination of the electron momentum at the primary interaction vertex as determined by the tracker, the energy of the corresponding ECAL cluster, and the energy sum of all bremsstrahlung photons spatially compatible with originating from the electron track. The energy of muons is obtained from the curvature of the corresponding track as determined using the tracker and the muon system. The energy of charged hadrons is determined from a combination of their momentum measured in the tracker and the matching ECAL and HCAL energy deposits, corrected for zero-suppression effects and for the response function of the calorimeters to hadronic showers. Finally, the energy of neutral hadrons is obtained from the corresponding corrected ECAL and HCAL energy deposits.

For each event, hadronic jets are clustered from these reconstructed particles
(particle-flow candidates) using the infrared- and collinear-safe anti-\kt algorithm~\cite{anti-kT}, as implemented in the \FASTJET package~\cite{fastjet}. The jet momentum is determined as the vectorial sum of all particle momenta in the jet, and is found from simulation to be, on average, within 5 to 10\% of the momentum of the particle-level jets reconstructed using stable particles (lifetime  $>$30\unit{ps}) excluding neutrinos, for jet $\pt>50\GeV$ and rapidity $\abs{y}<2.5$.
Additional proton-proton interactions within the same or nearby bunch crossings (pileup) can contribute additional tracks and calorimetric energy depositions, increasing the apparent jet momentum. To mitigate this effect, tracks identified as originating from pileup vertices 
are discarded and an offset correction~\cite{Cacciari:2007fd} is applied to correct for remaining contributions~\cite{CMS_JES_8TeV}. 
Additional selection criteria are applied to each jet to remove jets potentially dominated by instrumental effects or reconstruction failures~\cite{CMS-PAS-JME-16-003}. These criteria consist of the following conditions: the energy fraction of the jet carried by  neutral hadrons and photons should be less than $90\%$, the jet should have at least two constituents, and at least one of those should be a charged hadron. This set of criteria is more than $99\%$ efficient for genuine jets.

The missing transverse momentum vector (\ptvecmiss) is defined as the negative vector sum of the \pt of all reconstructed particle-flow objects in an event; its magnitude is denoted using {\ptmiss}. 
A set of algorithms is used to reject events with anomalous high-\ptmiss arising from a variety of reconstruction failures, detector malfunctions, or noncollision backgrounds; 
these algorithms are designed to identify 
more than 85--90\% of the spurious high-\ptmiss events 
with a misidentification rate of less than $0.1\%$~\cite{Sirunyan:2019kia}.

Jet energy corrections are derived using simulated \PYTHIA inclusive jet samples discussed in Sec.~\ref{MC}
so that the average measured transverse momentum of jets is the same as the corresponding particle-level jets.
This methodology is used to derive nominal jet energy correction factors only for AK4 and AK8 jets.
The nominal corrections for AK4 jets are used for AK1 to AK6 jets.
For larger jet sizes ($R>0.6$)
the nominal correction factors derived for AK8 jets are applied.
To account for the differences in the distance parameter, an extra correction factor ($C^{R}$) is determined in case of each distance parameter for the average pileup condition based purely on simulation and applied to the corresponding jets. A detailed discussion on the derivation of $C^{R}$ is made later in this section.
Measurements of the momentum balance in dijet, photon+jet, {\PZ}+jet, and multijet events are used to determine any residual differences between the jet energy scale in data and simulation, and appropriate corrections are made~\cite{CMS_JES_8TeV}.
The in situ techniques are based on the missing transverse momentum projection fraction method, which is fundamentally insensitive to jet size, and on particle-flow reconstruction, whose reliance on tracking and particle-flow hadron calibration further reduces differences in energy response between jets of different radii.
Residual corrections are derived using only AK4 jets 
and applied to jets of all the sizes.

The factor $C^{R}$ is derived in the following way. In simulated  \PYTHIA inclusive jet samples, the detector- and particle-level jets are required to be geometrically matched satisfying $\Delta R < (0.5 \times \textrm{jet size} ) $;
the ratio of the average detector-level jet \pt to the particle-level jet \pt is calculated as a function of the particle-level jet \pt for all the jet sizes 
and then used as an extra correction factor $C^{R}$ for both data and simulation.
The $C^{R}$ factors are also derived using simulated \HERWIGpp inclusive jet samples, so the jet energy response is calibrated to unity for all the jet sizes in \HERWIGpp samples as well.
The difference in $C^{R}$ factors derived using \PYTHIA and \HERWIGpp inclusive jet samples is used to estimate the systematic uncertainties in $C^{R}$ separately for each of the distance parameters. 
The $C^{R}$ factors are important to ensure that the jet energy resolution in simulation is properly corrected to match the data without changing the jet energy scale, and that \PYTHIA and \HERWIGpp are on equal footing with respect to jet energy scale and resolution when unfolding the data in Sec~\ref{sec_Unf}. 
The value of the $C^{R}$ factor ranges from 0.95 to 1.10 depending on the energy, rapidity, and size of the reconstructed jets; this correction is significant only for very small and very large jet sizes.
However, $C^{R}$ corrects the jet energy response for different jet sizes at the level of simulation only and no dedicated residual correction is derived for data as a function of jet size.
Nevertheless, it has been checked that the difference in average \pt between AK8 and AK4 jets pointing in the same direction in data is smaller than the difference of the same between \PYTHIA and \HERWIGpp  simulations.

\section{Event samples} \label{sec_Data}

\subsection{Collision data}

Proton-proton collision data collected by the CMS experiment during 2016, corresponding to an integrated luminosity of 35.9\fbinv,
are used for this analysis.  The data sample is collected using single-jet triggers, which select events containing at least one AK8 jet, formed from particle-flow candidates, with \pt exceeding one of the threshold values listed in Table~\ref{tab:hlt}.
Absolute trigger efficiency is measured with a tag-and-probe procedure~\cite{2011JInst611002C}
using the events having a back-to-back dijet topology, where the tag jet is matched to a single-jet trigger, and the efficiency is measured using the probe jet.

Because of limited bandwidth and storage space, only a fraction of the events
satisfying the triggering condition with lower thresholds are recorded. For this reason, in each jet-\pt bin, only the trigger that has the highest effective integrated luminosity and is also more than $99\%$ efficient is used. 

Offline, events are required to contain at least one jet with \pt above
that value for which the trigger is $99\%$ efficient.
These values are also used to define the \pt bins for the measurement.

\begin{table}[htbp]
\topcaption{Trigger \pt thresholds and effective integrated luminosity of the HLT triggers based on AK8 jets.
These triggers were not active during the initial part of data taking in 2016, 
thus the maximum integrated luminosity is less than 35.9\fbinv.
}
\centering
\begin{tabular}{p{3cm}  p{3cm}  p{3.5cm}}
\hline
Trigger \pt (GeV) \break threshold  & \pt (GeV) range \break for analysis & Effective integrated \break luminosity (fb$^{-1}$) \\
\hline
40 & 74--97 & 0.000050 \\
60 & 97--133 & 0.00033 \\ 
80 & 133--196 & 0.00104 \\
140 & 196--272 & 0.0105 \\
200 & 272--330 & 0.084 \\
260 & 330--395 & 0.517 \\
320 &  395--468 & 1.54 \\
400 &  468--548 & 4.68\\
450 &  548--$\infty$ & 33.4\\
\hline
\end{tabular}
\label{tab:hlt}
\end{table}

Similarly to other CMS publications~\cite{sigma7,sigma8}, 
we require that $ \ptmiss / \sum_{i} {\pt}^{i} < 0.3$, 
where the index $i$ runs over all particle-flow candidates in the event and $\sum_{i} {\pt}^{i}$ denotes the scalar sum of transverse momenta;
this rejects calorimeter noise and a part of electroweak backgrounds from the production of $\PW (\to$ lepton) $+$ jets, $\PZ (\to$ lepton) $+$ jets, and top quarks when the top quarks decay to final states with leptons.

\subsection{Simulated samples}
\label{MC}
The data are compared to predictions from several
different
MC generators, listed below.

The {{\PYTHIA}~v8.212}~\cite{pythia} generator computes matrix elements only for $2 \to 2$ Feynman diagrams at LO; the
missing orders in the perturbation series are approximated using \pt-ordered dipole showering. 
The {\PYTHIA} generator employs the empirical Lund string model to hadronize the partons. The NNPDF2.3~\cite{NNPDF} LO parton distribution function (PDF) set 
is used to describe the momentum fractions carried by the partons within the incoming protons; 
our UE model is the CUETP8M1 tune~\cite{CUETP8M1_CMS} (CMS Underlying Event Tune for {\PYTHIA}8 based on Monash~\cite{Monash}), which was derived by tuning the model parameters using minimum bias data collected by the CMS Collaboration.

The {{\HERWIGpp}~v2.7.1}~\cite{herwig} generator also calculates only $2 \to 2$ scatterings,
but has a different fragmentation and hadronization model than {\PYTHIA}. It employs angular-ordered showers to radiate the partons and a cluster model to produce the hadrons. The NNPDF3.0 LO PDF set is used, and the UE modeling is described by the CUETHppS1 tune~\cite{CUETP8M1_CMS}.
 
The {\MADGRAPH ({\MGvATNLO}~V5~2.2.2)}~\cite{madgraph1} generator provides calculations of matrix elements with up to four outgoing partons in the final state at LO. The partons
  are showered and hadronized with {\PYTHIA} combined with {\MADGRAPH}, using the MLM merging scheme~\cite{madgraph2}. The NNPDF3.0 NLO PDF set and the CUETP8M1 UE tune are used here as well. 

The {{\POWHEG}~v2}~\cite{Powheg_ref,Alioli:2010xa} generator computes the dijet production cross section at NLO in \pQCD. Successive parton showering, hadronization, and UE modeling is performed either using  \PYTHIA with the CUETP8M1 tune (referred to as PH$+$P8 in the figures) or \HERWIGpp with the CUETHppS1 and EE5C~\cite{Seymour:2013qka}  tunes (referred to as PH$+$Herwig in the figures).
  The NNPDF3.0 NLO PDF set is used for \POWHEG as well, and the value of the $h_{\text{damp}}$ parameter in \POWHEG is 250\GeV.

The {{\HERWIG}~7.1.1}~\cite{Bellm:herwig7} generator, used with the NNPDF3.1 NNLO PDF set with $\alpS \ (m_{\PZ})=0.118$, also evaluates the matrix elements for dijet production at NLO, and is matched to the
{\HERWIGS} parton shower using the FxFx~\cite{Frederix:2012ps} jet merging method. 
The CH2 tune is used to model UE.  This prediction is referred to as {\HERWIGS}.

Fixed-order predictions for dijet production at NLO are computed using \NLOJET within the framework of the \FASTNLO package. To account for the effects of hadronization,
an additional correction factor is used, which will be discussed in Section~\ref{sec_NPCor}. 
This prediction is referred to as NLO$ \otimes $NP in the figures.
Predictions from \NLOJET are obtained using the CT14NLO PDF set. 

Recently, a prediction for single-inclusive jet production
using joint resummation in the threshold energy in the small-R limit has
been computed at
next-to-leading logarithmic (NLL) accuracy in the framework of Soft Collinear Effective Theory in Refs.~\cite{Liu:2017pbb,Liu:2018ktv};
the  CT14NLO PDF set is also used for this prediction, which is referred to as (NLO+NLL).
This prediction is compared with the measurements reported in this paper.

\section{Measurement of cross sections and cross section ratios}
 
The inclusive jet cross section is calculated as
\begin{equation}
\frac{{\rd}^{2}\sigma}{\rd\pt \rd y} = \frac{1}{ \epsilon  \mathcal{L_{\text{int}}}   } \frac{N_{\text{jets}}}{\Delta \pt \Delta y}  , 
\label{pteq}
\end{equation}
where $N_{\text{jets}}$ is the number of jets in a \pt and $y$ bin, $\mathcal{L_{\text{int}}}$ is the integrated luminosity of the data set, $ \epsilon$ is the product of trigger and event selection efficiencies, and $\Delta \pt$ and $\Delta y$ are the bin widths in \pt and $y$. The widths of \pt bins are proportional to the jet energy resolution and increase with jet \pt.
The ratios of cross sections for the different jet sizes with respect to AK4 jets is
calculated as
the bin-by-bin quotient of the cross sections of AK\textit{n} (\textit{n}$=1,2,...,12$)
and AK4 jets respectively;
in the ratios,
all the terms in Eq.~\eqref{pteq} except ${N_{\text{jets}}}$ and $\epsilon$ cancel.

\subsection{Unfolding} \label{sec_Unf}

To correct for detector inefficiencies and resolution,
a number of methods available in the 
\mbox{\sc RooUnfold} package~\cite{Adye:2011gm}
are used to unfold the jet \pt spectra.

The nominal choice of unfolding technique in this paper is the D'Agostini unfolding~\cite{bayes} with early stopping.
Up to 5--8 iterations are used depending on jet size and rapidity region.  
An alternative method is singular value decomposition (SVD)~\cite{SVD}.
A third method is called bin-by-bin~\cite{Schmitt:2016orm}, which multiplies the particle-level spectra by the ratio between the detector-level spectra in data and simulation.

The SVD and bin-by-bin techniques are used to cross-check the result of unfolding with the D'Agostini unfolding.
As an additional cross-check, unfolding is also performed using a
$\chi^2$ minimization without regularization using the \textsc{TUnfold} package~\cite{tunfold}.

\begin{figure}[htbp]
\centering
\includegraphics[width=0.45\textwidth]{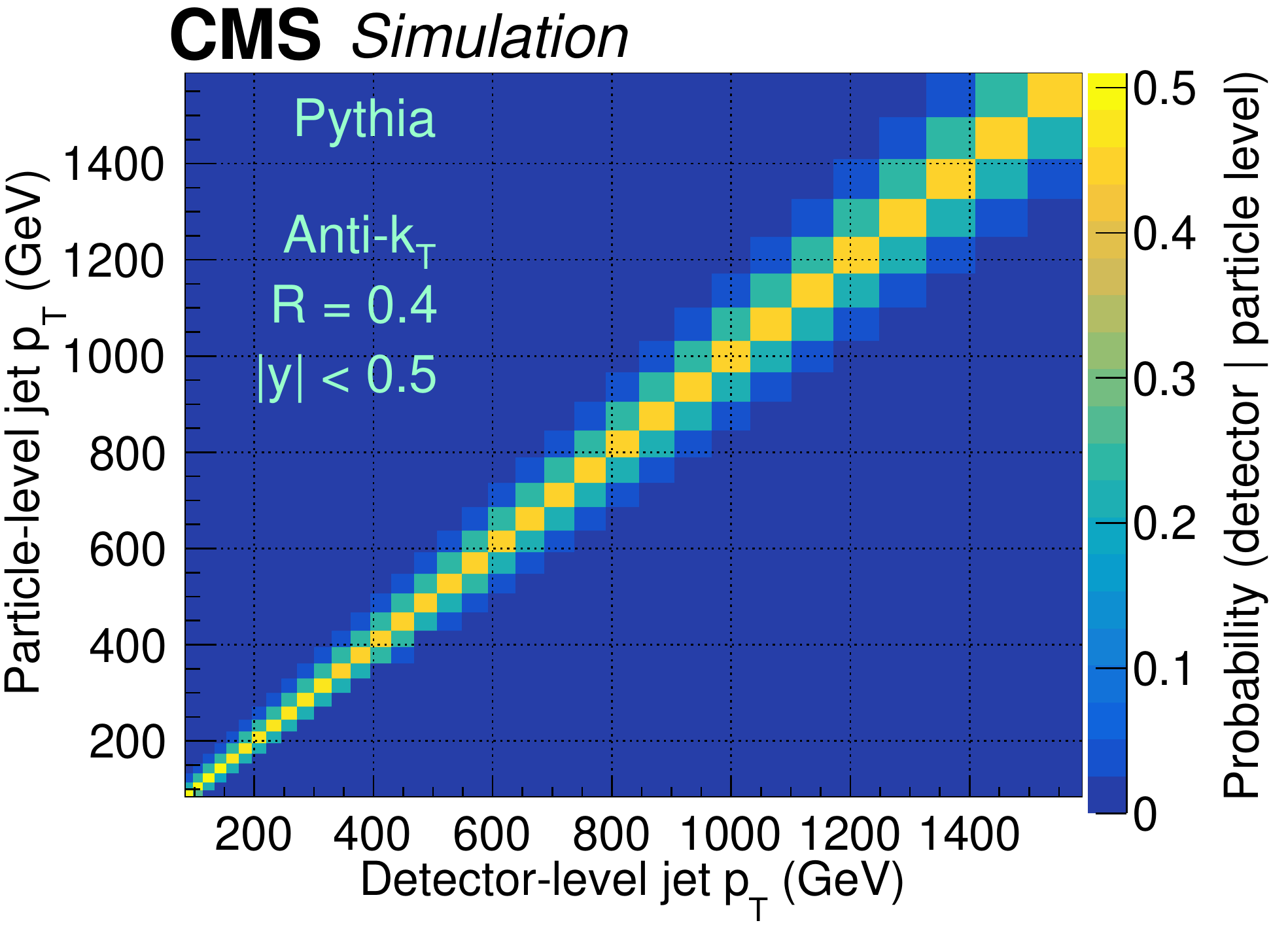}
\includegraphics[width=0.45\textwidth]{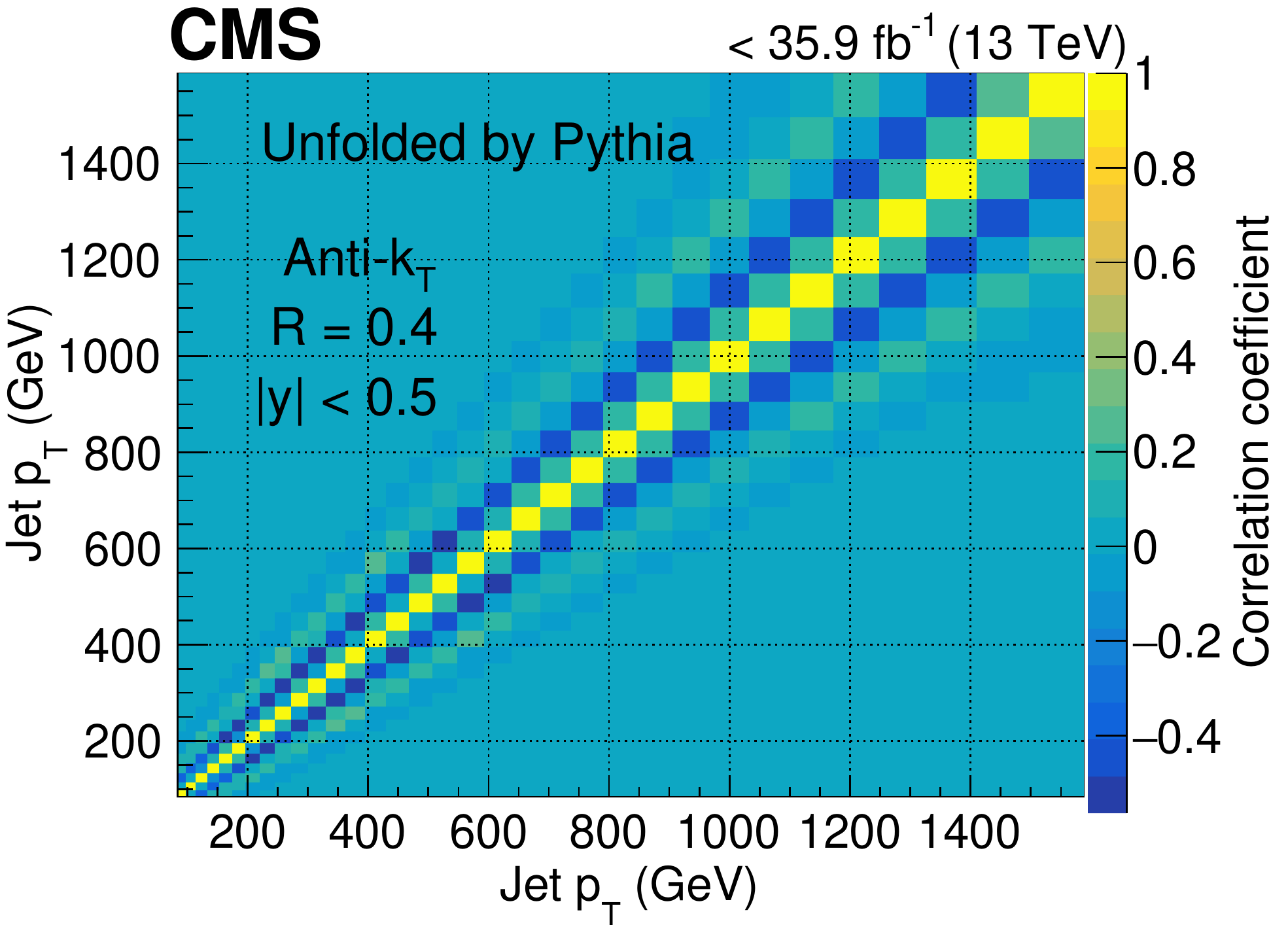}
\caption{Response matrix constructed from a simulation of a sample generated using {\PYTHIA}, for AK4 jets in the $\abs{y}<0.5$ bin (left). A correlation matrix generated after data is unfolded by the D'Agostini unfolding  using {\PYTHIA} simulation for AK4 jets (right).}
\label{fig:RM1}
\end{figure}

Response matrices between \pt spectra of detector-level and generator-level jets are obtained
by one-to-one matching of the nearest  detector- and particle-level jets, excluding matches with
$\Delta R> ( 0.5 \times \text{jet size} )$, where $\Delta R$ denotes the distance between detector- and particle-level jets in the rapidity-azimuth plane. 
This criterion leads to almost 100\% matching efficiency between the detector-level and the particle-level jets.
Response matrices are constructed, for all rapidity and jet sizes, from the CMS detector simulation based on \GEANTfour ~\cite{Agostinelli:2002hh}
using simulated samples from three MC event generators,  {\PYTHIA}, \HERWIGpp, and \MADGRAPH. 
For the particle-level results, response matrices based on the {\PYTHIA} simulation are used for the unfolding. 
The response matrix for AK4 jets in the first rapidity region for the {\PYTHIA} sample is shown in Fig.~\ref{fig:RM1}.
The response matrix is diagonal, which shows that unfolding works well.

For both the D'Agostini and SVD unfolding techniques, the nearest neighbor \pt bins are correlated,
and the next-to-nearest bins are anti-correlated (right plot in Fig.~\ref{fig:RM1}
for AK4 jets with the D'Agostini unfolding). Next-to-next-to-nearest bins are again correlated.

Several cross-checks are made regarding the unfolding.
To investigate possible bias due to the choice of MC generator used
to construct the response matrices, event samples are generated using three different generators: {\PYTHIA}, \HERWIGpp, and \MADGRAPH, followed by the detector simulation
whose output is scaled and smeared independently for each generator to match the energy scale and resolution of jets in data.
Detector-level distributions from
each of the samples are unfolded using these three response matrices, 
and the unfolded distributions are compared to the corresponding particle-level distributions.
No evidence for significant bias is observed. 
Similarly, the data are unfolded using response matrices from these three simulated samples; 
the differences among the unfolded spectra are within systematic uncertainties corresponding to the correction factor $C^R$. 
The same conclusion holds when comparing the unfolded distributions obtained using different unfolding techniques, such as D'Agostini, SVD, bin-by-bin, and $\chi^2$ minimization.

\subsection{Nonperturbative corrections for fixed-order calculations}  \label{sec_NPCor}

Fixed-order NLO calculations yield predictions for the partonic fields,
but in experimental measurements, jets are composed of hadrons. To evolve the parton-level prediction to the hadron level, NP corrections are calculated and applied.  Although generators such as \PYTHIA and \HERWIG come with MC-based phenomenological simulation of these processes, \NLOJET does not. 
The impact of NP on the \NLOJET prediction is approximated as a multiplicative correction factor as follows.
The NP correction is the ratio of an observable from a generator, which includes NP effects
with hadronization and multiple parton interaction (MPI) processes switched on, to the same observable obtained from the same generator without NP effects, i.e., by switching off  hadronization and MPI processes.

Simulated \POWHEG{}$+$\PYTHIA (CUETP8M1 tune) 
and \POWHEG{}$+$\HERWIGpp
(EE5C and CUETHppS1 tunes) samples
are used to compute NP factors for all the jet sizes in all the \pt and rapidity bins. 
The average NP correction obtained from the \POWHEG{}$+$\PYTHIA and \POWHEG{}$+$\HERWIGpp (EE5C) samples is defined as the final NP correction, and
the envelope of the differences
is taken as its uncertainty.

\begin{figure}[htbp]
\centering
\includegraphics[width=0.45\textwidth]{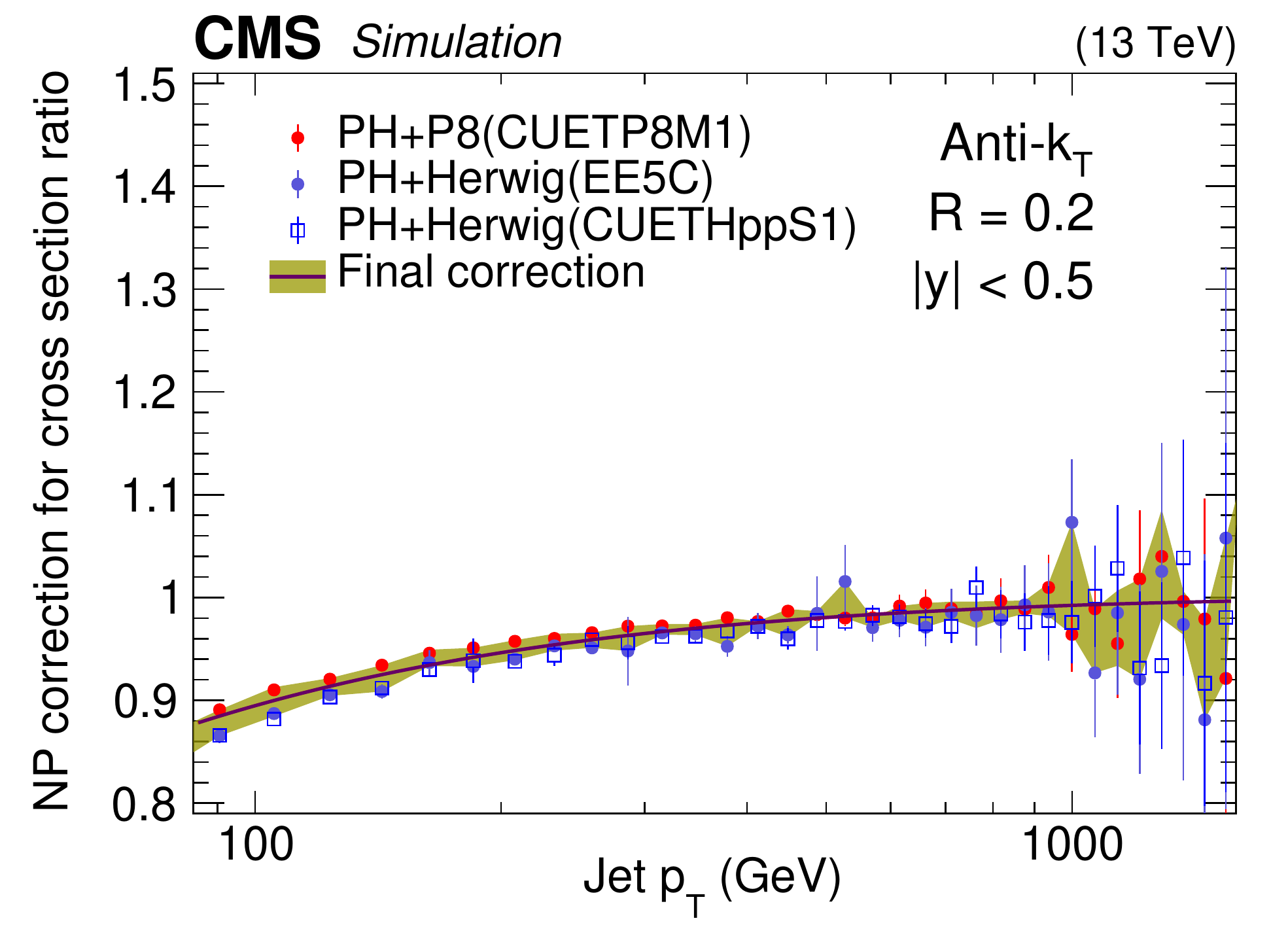}
\includegraphics[width=0.45\textwidth]{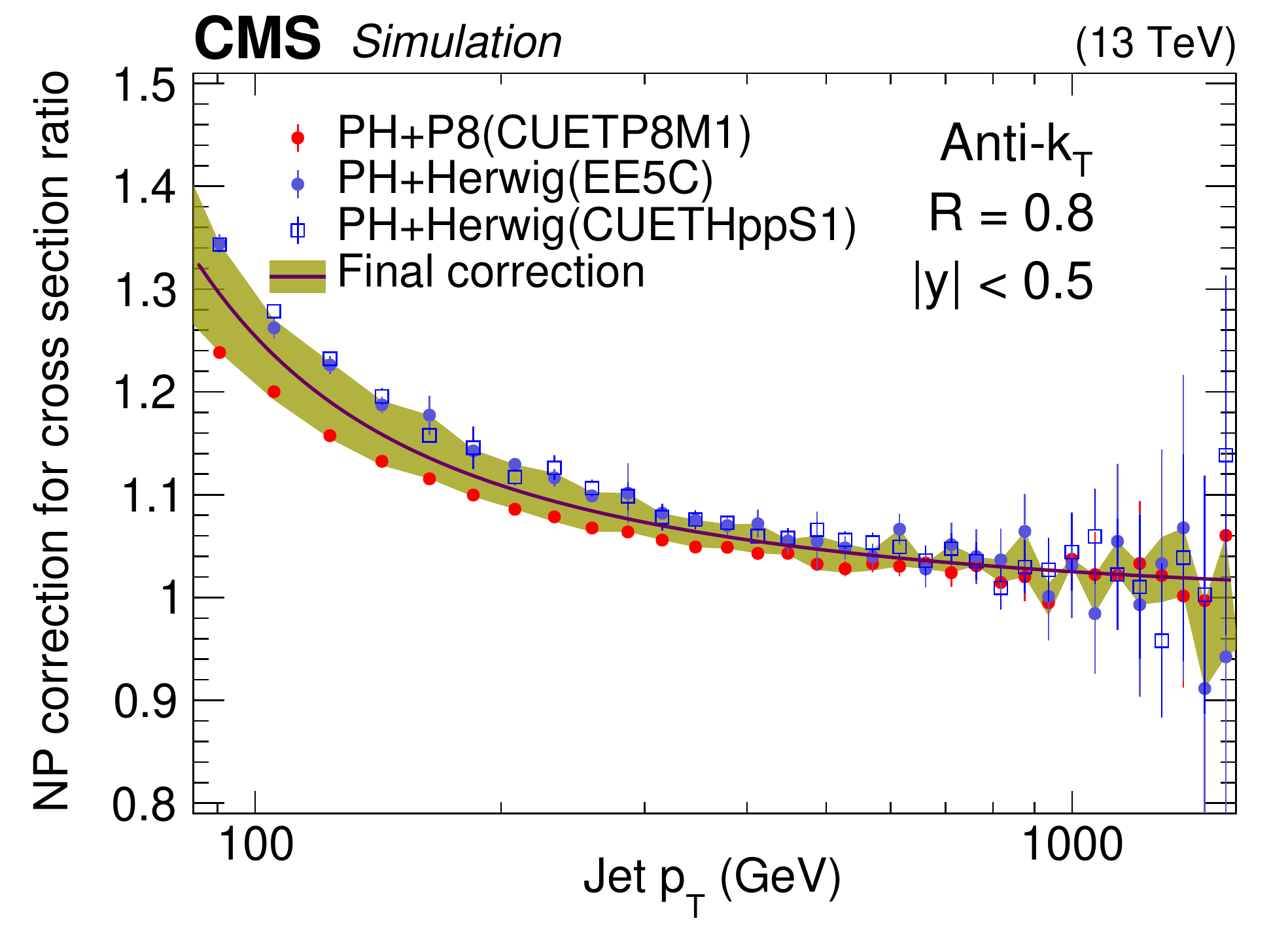}
\caption{Nonperturbative correction factor for the cross section ratio of inclusive AK2 (left) and AK8 jets (right) with respect to the AK4 jets in the rapidity bin $\abs{y}<0.5$. 
Vertical error bars represent the statistical uncertainty of the NP correction for different predictions.}
\label{fig:npcor}
\end{figure}

Figure~\ref{fig:npcor} depicts the NP corrections for the cross section ratio of the AK2 and AK8 jets 
with respect to the AK4 jets.
Hadronization corrections are larger
for smaller jet sizes, and MPI introduces a larger correction for large-$R$ jets.
Because both hadronization and MPI are important for low-\pt jets,
the NP correction is also significant in the low-\pt portion of phase space;
in the high-\pt region, the NP correction factor approaches 1. For AK4 jets,
the corrections for hadronization and MPI almost cancel,
and the resulting NP correction is close to unity throughout the \pt range.
At around $\pt=85\GeV$, the correction
goes down to 0.8 for AK2 jets, and it goes up to 1.25 for AK8 jets.

\section{Experimental uncertainties in the measurement}

Multiple sources of uncertainty affect the precision of the measurement: statistical, jet energy scale (JES) uncertainties,
jet energy resolution (JER) uncertainties, and uncertainties in the pileup condition. 
We also include systematic uncertainties corresponding to 
the use of JES corrections derived for one $R$
along with the $R$-dependent $C^{R}$ factor
on jets formed using another $R$.

To estimate the statistical uncertainty in data, the
\textit{jackknife resampling}~\cite{Efron:526679} method is used.
In this technique, ten different data samples, each containing 90\% events of the full data sample, are constructed such that the removed 10\% of the events are complementary for each subsample.
These subsamples are chosen in such a way that they correspond to very similar phase space regions.
The statistical uncertainty is the standard deviation of the
ten distributions multiplied by $\sqrt{9}=3$. The resulting
statistical uncertainty is roughly $<$1\% for jet  $\pt < 1$\TeV,
and increases at high jet \pt. 
A similar procedure is followed to estimate the statistical uncertainty
due to the response matrices used for unfolding.
Here also, ten subsets of the simulated sample are considered, each
with a  nonoverlapping 10\% of events removed.
The distributions in data are unfolded using each subsample, and the standard deviation of ten unfolded distributions, multiplied by a factor of 3, is the statistical uncertainty due to the response matrices; in this case the statistical uncertainty is roughly 0.5--1.0\% for the cross section ratio throughout the \pt range.

The jet energy scale corrections have a number of uncertainties
corresponding to the techniques used and the amount of pileup.
The JES has an uncertainty of about 1--2\% in the central region~\cite{CMS-DP-2018-028}.
The uncertainty is larger in the forward region and at low jet \pt. 
To evaluate the uncertainty in the measurement of the cross section ratio,
the JES is varied upwards and downwards by the uncertainties corresponding to
different sources.
The difference in the unfolded cross section ratios
using the nominal and varied JES is the uncertainty.
Twenty-seven different sources of JES uncertainty are considered
individually and added in quadrature.
The uncertainty because of JES 
is very similar for all the jet sizes, except for the pileup component.
The uncertainties mostly cancel out in the ratio, but there is a small residual,
which is about 0.5--1.0\% for $\abs{y}\leq2.0$ up to 1\TeV of jet \pt and goes up to 2\% for very high jet \pt.

To estimate the uncertainty in the ratio of cross sections with respect to that of AK4 jets because of
using JES corrections derived for one value of $R$ with jets from other values of $R$
and then applying the $C^{R}$ factor, 
the standard calibration factors from the AK8 jets are applied to AK1 to AK6 jets, 
and, for jets of other sizes, the calibration factors for the AK4 jets are used.
The $C^{R}$ factors for jets of all sizes are derived for this scenario. 
The systematic uncertainties in the inclusive jet cross section ratios are evaluated using the difference between the results obtained by these two procedures. 
The uncertainty coming from the $C^{R}$ correction is more significant for larger jet sizes.

The $C^{R}$ calibration factors
are derived using both {\PYTHIA} and {\HERWIGpp} simulations
as a function of jet \pt in different rapidity bins for all jet sizes.
The difference in the resulting $C^{R}$ corrections
is an `$R$-dependent' uncertainty,
and it is defined such that it vanishes for AK4 jets,
which is used as the reference. 

\begin{figure}[htbp]
\centering
\includegraphics[width=0.7\textwidth]{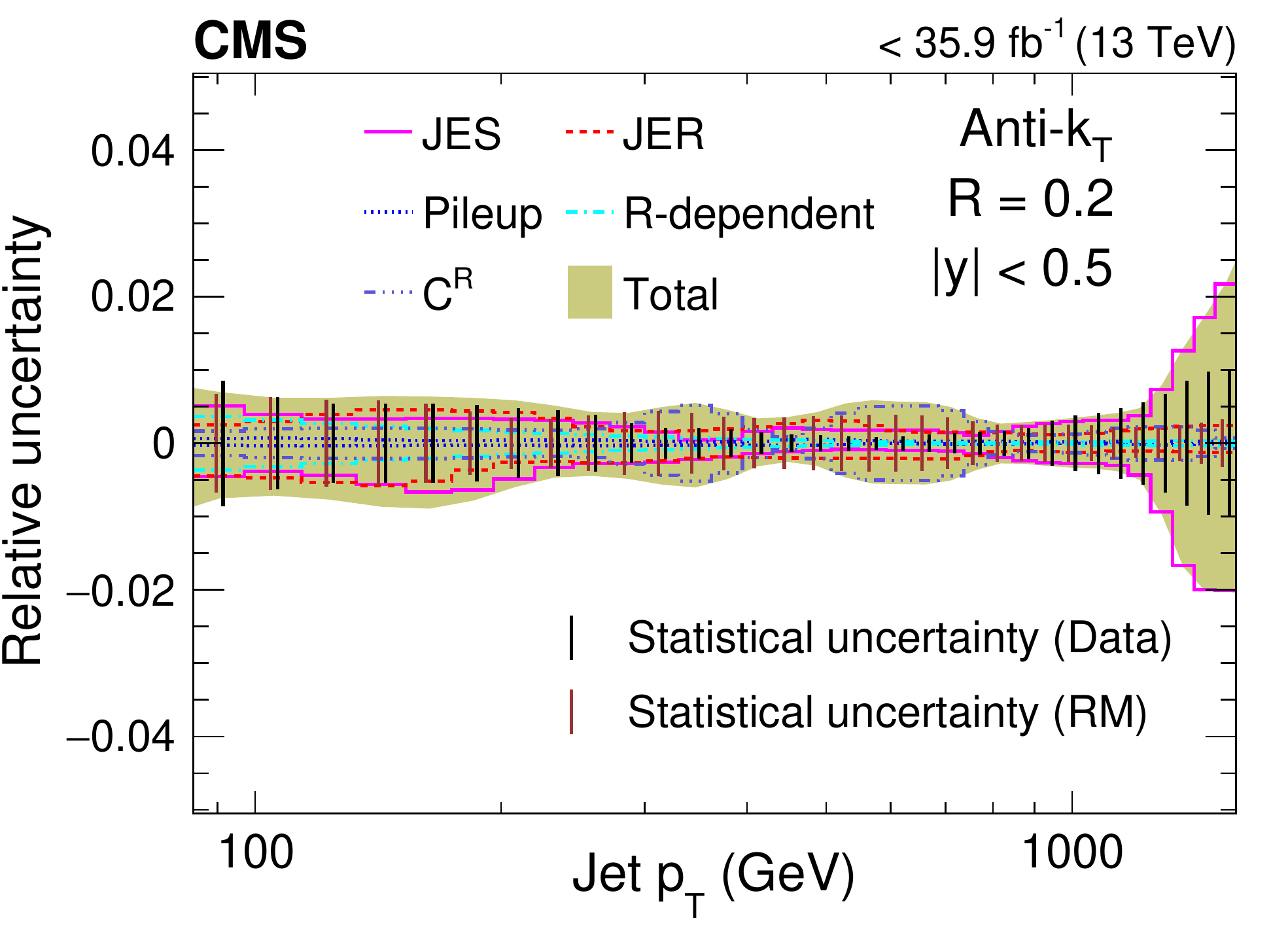}
\includegraphics[width=0.7\textwidth]{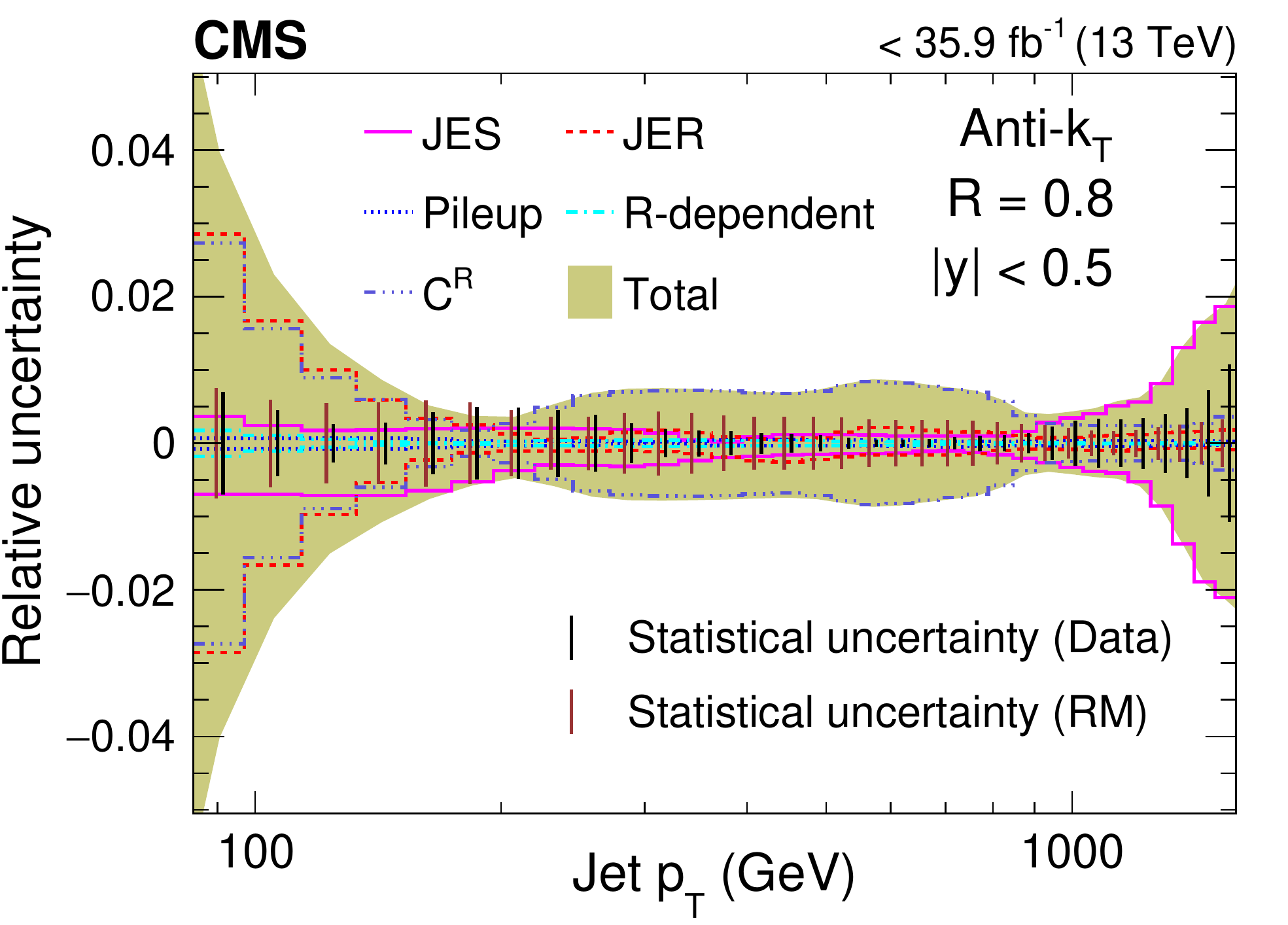}
\caption{Total uncertainty (relative) from experimental sources for the ratio of cross section of inclusive jets of size 0.2 (top) and 0.8 (bottom) with respect to that of AK4 jets in the rapidity bin $\abs{y}<0.5$. Statistical uncertainties are also overlaid as vertical black (red) bars for data (response matrices, RM, in simulation).}
\label{fig:unc_tot_rat}
\end{figure}

The JER and its associated uncertainty are obtained from a dijet balance technique~\cite{CMS-DP-2018-028}.
The JER in data is worse than in simulation.
To match the JER in data and simulation, a spreading is added to the jets in simulation.
Here also, as in the case for the JES,
cross section ratios are obtained using upward and downward variations of the energy resolution factors for simulation while unfolding the data. 
The difference with respect to the nominal unfolding is used as an estimate of the uncertainty.
The uncertainty due to JER is more important for large-$R$ jets
at low \pt . The uncertainty also grows in regions of larger rapidities.

To match the pileup conditions in data and in MC simulation,
pileup profile
weighting is performed for the simulated samples.
The weighting factors depend on the
total inelastic cross section;
we vary its nominal value of  67.5\unit{mb}~\cite{CMS_MB} up and down by its uncertainty of $2.6\%$ when reconstructing the response matrices, and take the difference in the unfolded data as the uncertainty. 
This source of systematic uncertainty is larger at low \pt for large jet sizes, although its absolute value is small.

The uncertainties from different experimental sources are added in quadrature, 
and the total uncertainty is shown in Fig.~\ref{fig:unc_tot_rat} for the cross section ratios of the AK2 and AK8 jets with respect to the AK4 jets.
 
In the cross section ratio, many of the systematic uncertainties almost cancel,
so the final uncertainty is small. 
The statistical component of the uncertainty is also shown in the same figure. 

The  experimental systematic uncertainty at low \pt and large $R$ is dominated  by
the pileup uncertainty. The JER uncertainty is also larger there because of additional spreading caused by pileup. At intermediate \pt, the uncertainty is dominated by the $C^{R}$ uncertainty;
at high \pt the JES dominates the experimental uncertainty
because the cross sections fall very steeply and event counts are small at high jet \pt .
The sizes of the 
statistical uncertainties are similar to those of the total systematic uncertainties
and are dominated by data at high \pt and by the uncertainty in the response matrix because of the number of MC events
at intermediate \pt. At low \pt, the data have similar statistical uncertainties as the simulated sample,
since the corresponding triggers are prescaled.

Another source of uncertainty, which is relevant only for jets with $R>0.8$, is the uncertainty in the trigger efficiency correction. The AK8 single-jet triggers are not fully efficient for larger jet sizes near the trigger turn-on points for AK8 jets; an efficiency correction is applied for those jet sizes following Eq.~\eqref{pteq}. The difference in the absolute value of the trigger efficiency
from the curve used to model the variation of trigger efficiency as a function of jet \pt
is the uncertainty.
The size of this uncertainty is 0.5--1.0\% throughout the \pt range.

\section{Theoretical uncertainties} \label{sec_sys_th}

Apart from the systematic uncertainties due to experimental sources, theoretical calculations and generators have uncertainties in their predictions for the cross section ratio. For the fixed-order predictions, the contributing factors include the choice of renormalization and factorization scales (scale), the PDF uncertainty (PDF), the uncertainty from {\alpS}, and the uncertainty due to the NP corrections (NP correction).

In the matrix element computation, the coupling (\alpS for QCD) is evaluated at an energy scale known  as the renormalization scale ($\mu_{\mathrm{R}}$).
Another scale is chosen to compute the PDF, in order to resum initial-state radiation below that scale, called the factorization scale ($\mu_{\mathrm{F}}$). 
For the fixed-order calculations, both are set equal to the \pt of individual jets.
The scale uncertainty 
is evaluated using the following combinations of factors for ($\mu_{\mathrm{R}}$, $\mu_{\mathrm{F}}$) : (2, 1), (1, 2), (0.5, 1), (1, 0.5), (2, 2), (0.5, 0.5). The envelope of the variations is the scale uncertainty in the prediction.
Scale variations in the numerator and the denominator of the ratio of cross sections are fully correlated since the underlying parton configuration is the same.
This is one of the largest sources of theoretical uncertainties.

The PDFs are determined using data from several experiments.
The PDFs therefore have uncertainties from the experimental measurements,
modeling, and parameterization assumptions.
The resulting uncertainty is calculated
according to the prescription of CT14~\cite{CT14} at the $90\%$ confidence level and then scaled to the $68.3\%$ confidence level.
The PDF uncertainty is independent of jet size within statistical uncertainties, and thus cancels in the ratios.

The cross section measurement for inclusive jets depends on the value of \alpS.
In the \NLOJET prediction, its value (0.118) is varied by $\pm 0.001$. 
The uncertainty 
is taken as the
difference between the results  with varied and
nominal values of \alpS
and this difference is scaled to correspond to
$\Delta \alpS \simeq 0.0015$, as recommended in Ref.~\cite{PDF4LHC}.
For the jet cross section ratio, the uncertainty due to the \alpS variation
in the numerator and denominator cancels.

As mentioned in Section~\ref{sec_NPCor}, the envelope of the differences between
the NP correction factors obtained using different
parton showering algorithms to determine the NP correction
is the uncertainty in the NP correction.
The uncertainty is significant only at low {\pt}.

All these uncertainties are added in quadrature, and
are collectively referred as the theoretical uncertainty in what follows. 

The correlation between the experimental and the theoretical uncertainties is not studied.

\section{Results}

\subsection{Comparison of ratio of cross sections} \label{sec_xsec_rat_gen}

\begin{figure}[htbp]
\centering
\includegraphics[width=0.8\textwidth]{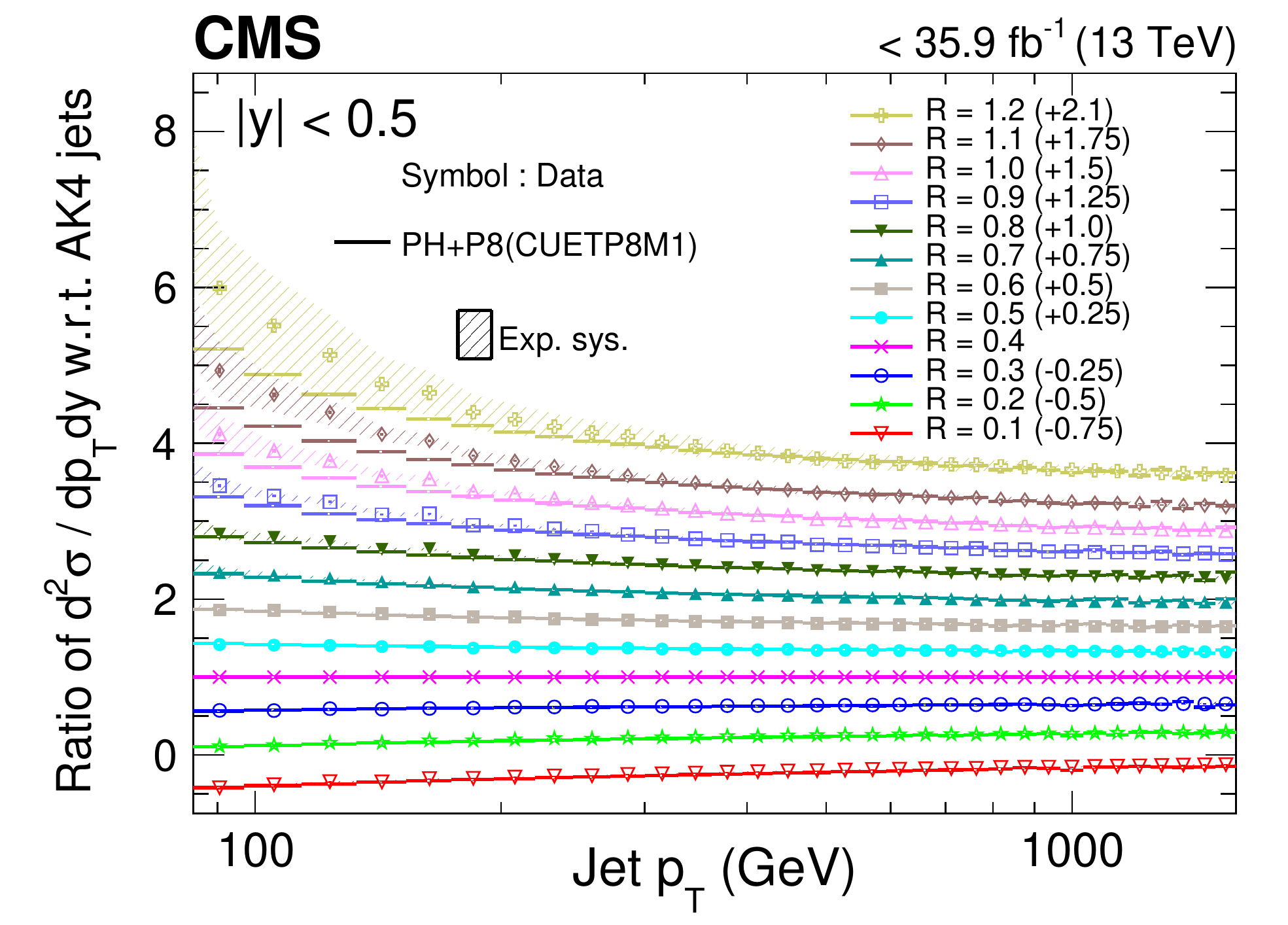}
\caption{Comparison of the ratio of the differential cross sections of jets of different sizes with respect to that of AK4 jets from data and from NLO predictions using \POWHEG{}$+$\PYTHIA (CUETP8M1 tune) in the region $\abs{y}<0.5$. Colored symbols indicate data and colored lines represent prediction from simulation. Offsets by the amount written in the parentheses have been added to the corresponding data points to separate the results for different jet sizes.}
\label{fig:genptrat2}
\end{figure}

The ratios of cross sections with respect to the AK4 jets are shown in Fig.~\ref{fig:genptrat2} in the central region ($\abs{y}<0.5$) 
for all the jet sizes
using unfolded data and the prediction from the NLO MC generator \POWHEG with \PYTHIA parton showering;
they are offset by fixed quantities for clarity.

The NLO \POWHEG generator, interfaced with the parton showering model, describes the data well at moderate values of jet size, but there is a deviation at low \pt for very large values of jet size. 

\begin{figure}[htbp]
\begin{center}
\includegraphics[width=0.75\textwidth]{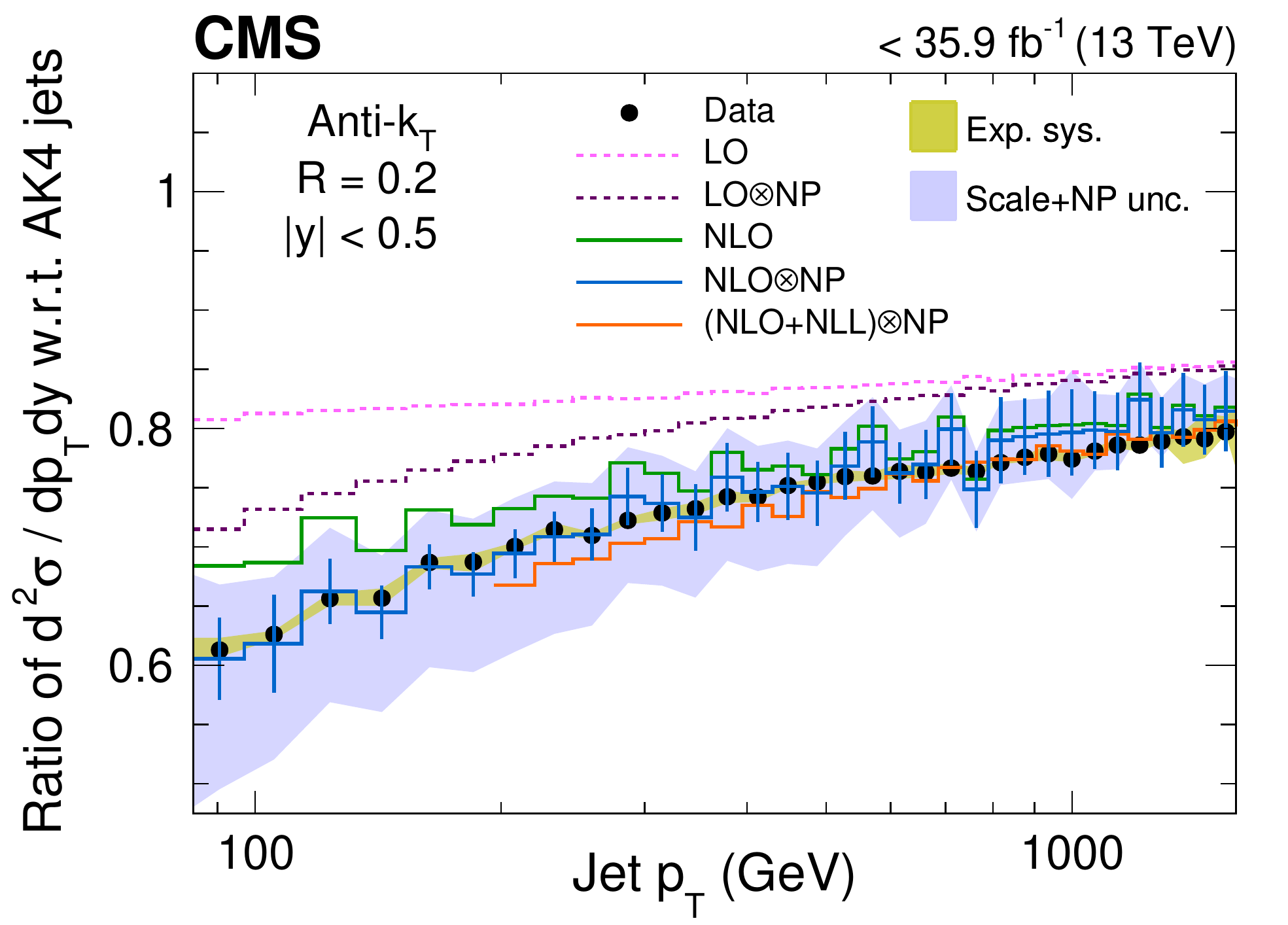}
\includegraphics[width=0.75\textwidth]{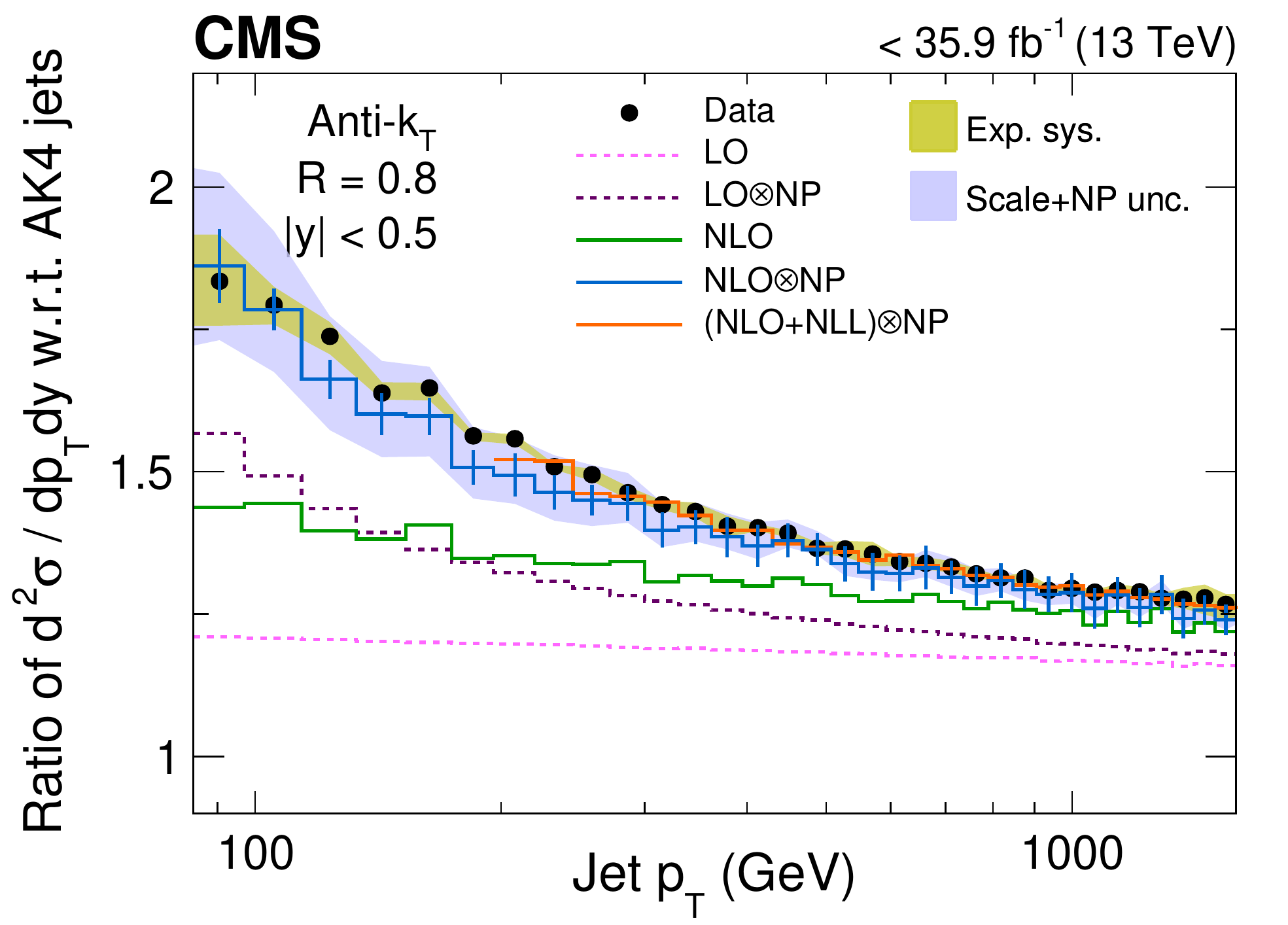}
\end{center}
\caption{Comparison of the ratios of differential cross sections for the AK2 (upper) and AK8 (lower) jets with respect to that of AK4 jets from data and \pQCD predictions using \NLOJET in the region $\abs{y}<0.5$. Black symbols indicate data and colored lines represent \pQCD predictions. Statistical uncertainties are shown as vertical bars for the data and the NLO$\otimes$NP prediction. The yellowish olive region around data represents the experimental systematic uncertainty whereas the region shaded in light blue color around NLO$\otimes$NP prediction shows the theoretical uncertainty in the prediction.}
\label{fig:genptrat4}
\end{figure}

The ratios of the cross sections of inclusive AK2 and AK8 jets with respect to those of AK4 jets are computed at LO and NLO
in \pQCD, following Eq.~\eqref{rad_diff}, with \NLOJET for the most central region ($\abs{y}<0.5$). 
The comparison with data is shown in Fig.~\ref{fig:genptrat4}.
Both the LO and NLO predictions are systematically below data for AK8 jets 
and above data for AK2 jets.
The NP correction is essential to describe the trend in data below medium jet \pt values.
Also, the NLO calculation improves data-theory agreement significantly over LO, bringing data and theoretical prediction into agreement within statistical and systematic uncertainties at $\pt>1000\GeV$ for both AK2 and AK8 jets.
Resummed calculations bring the theoretical prediction even closer to the data, especially for AK8 jets. 
The uncertainty corresponding to the resummed calculations is within $5\%$ for cross section ratio, and is not shown here to avoid congestion in the figure. 

\subsection{Variation of the ratio of cross sections with jet size} \label{sec_xsec_int_gen}

\begin{figure}[htbp]
\centering
\includegraphics[width=0.33\textwidth]{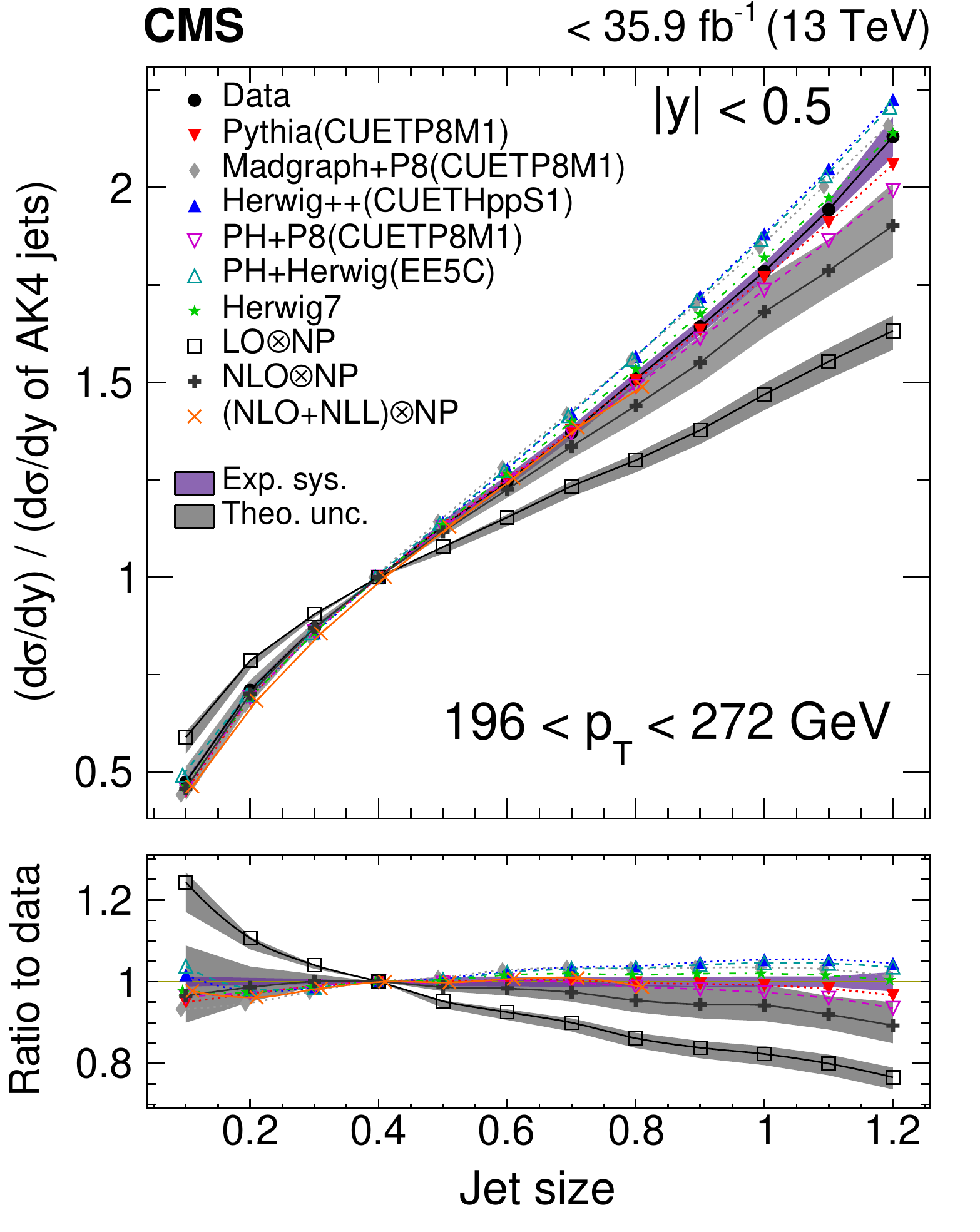}
\includegraphics[width=0.33\textwidth]{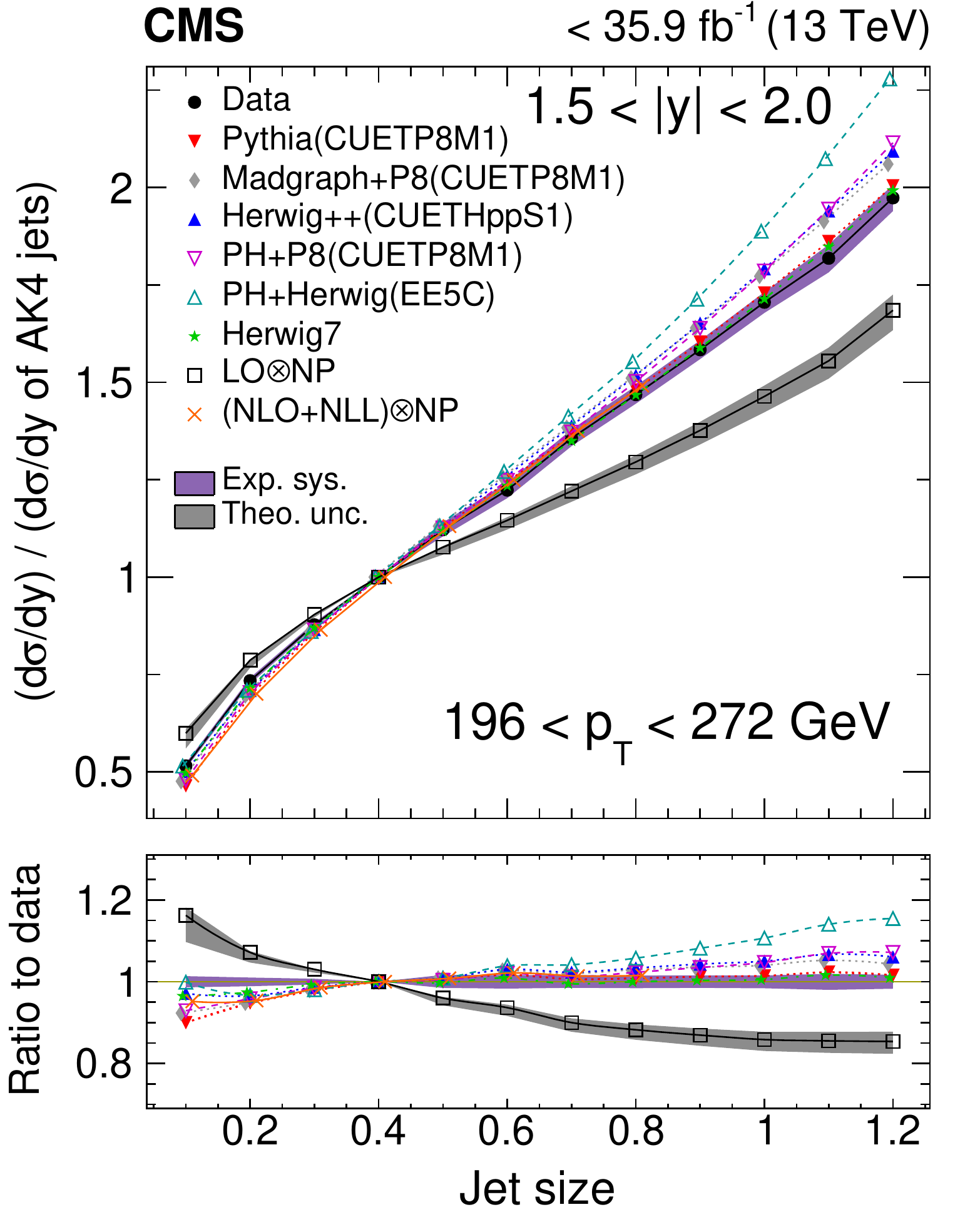}
\includegraphics[width=0.33\textwidth]{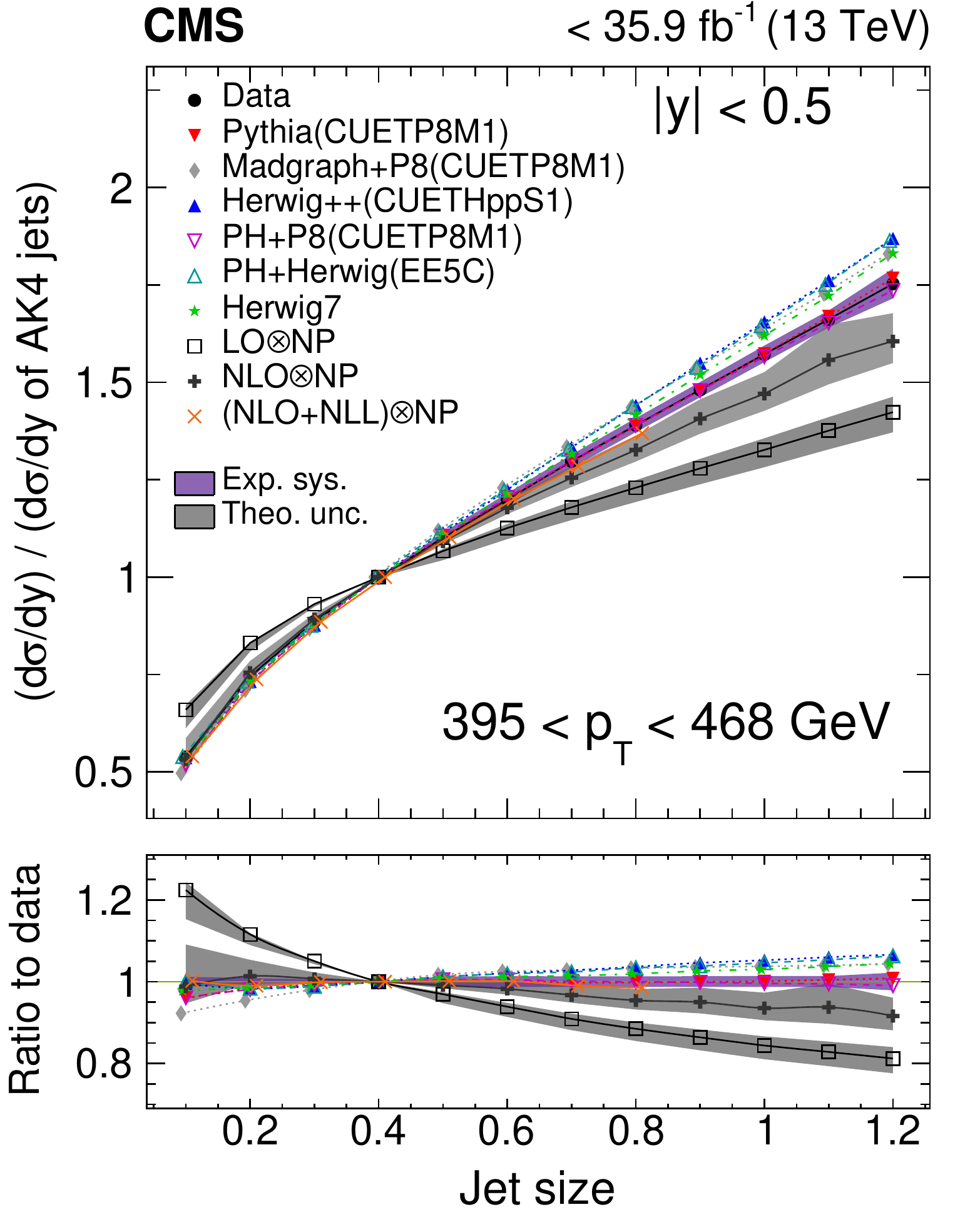}
\includegraphics[width=0.33\textwidth]{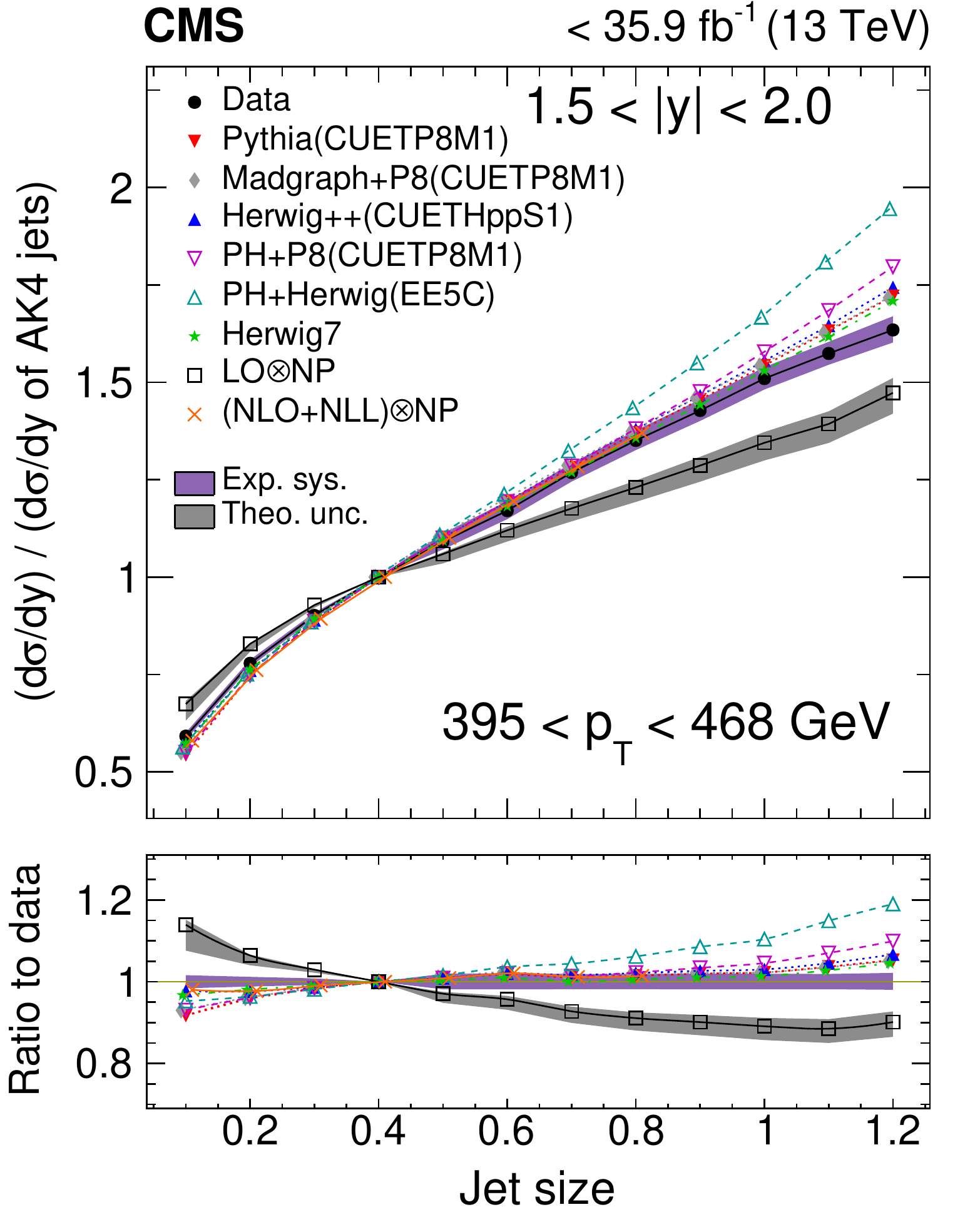}
\includegraphics[width=0.33\textwidth]{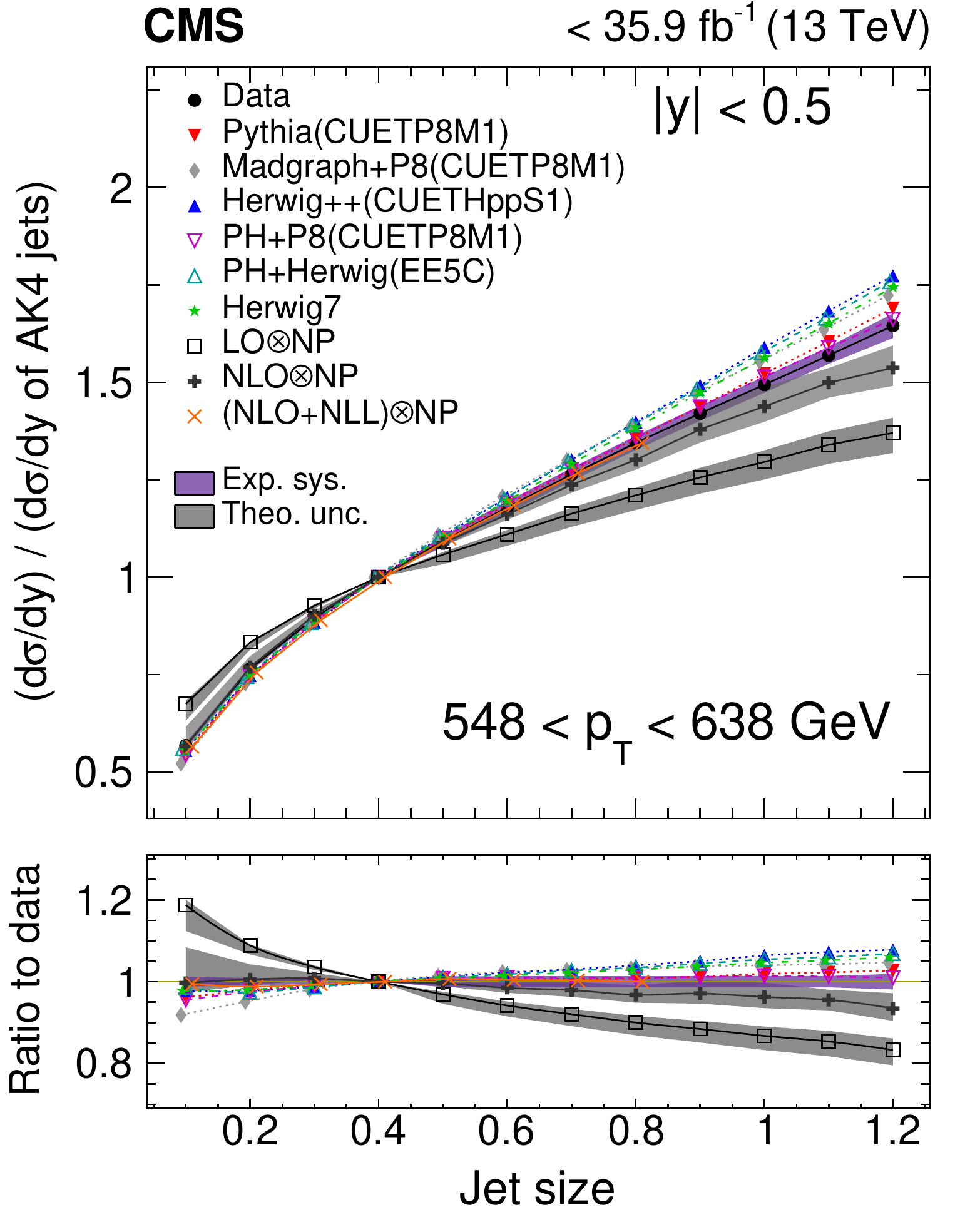}
\includegraphics[width=0.33\textwidth]{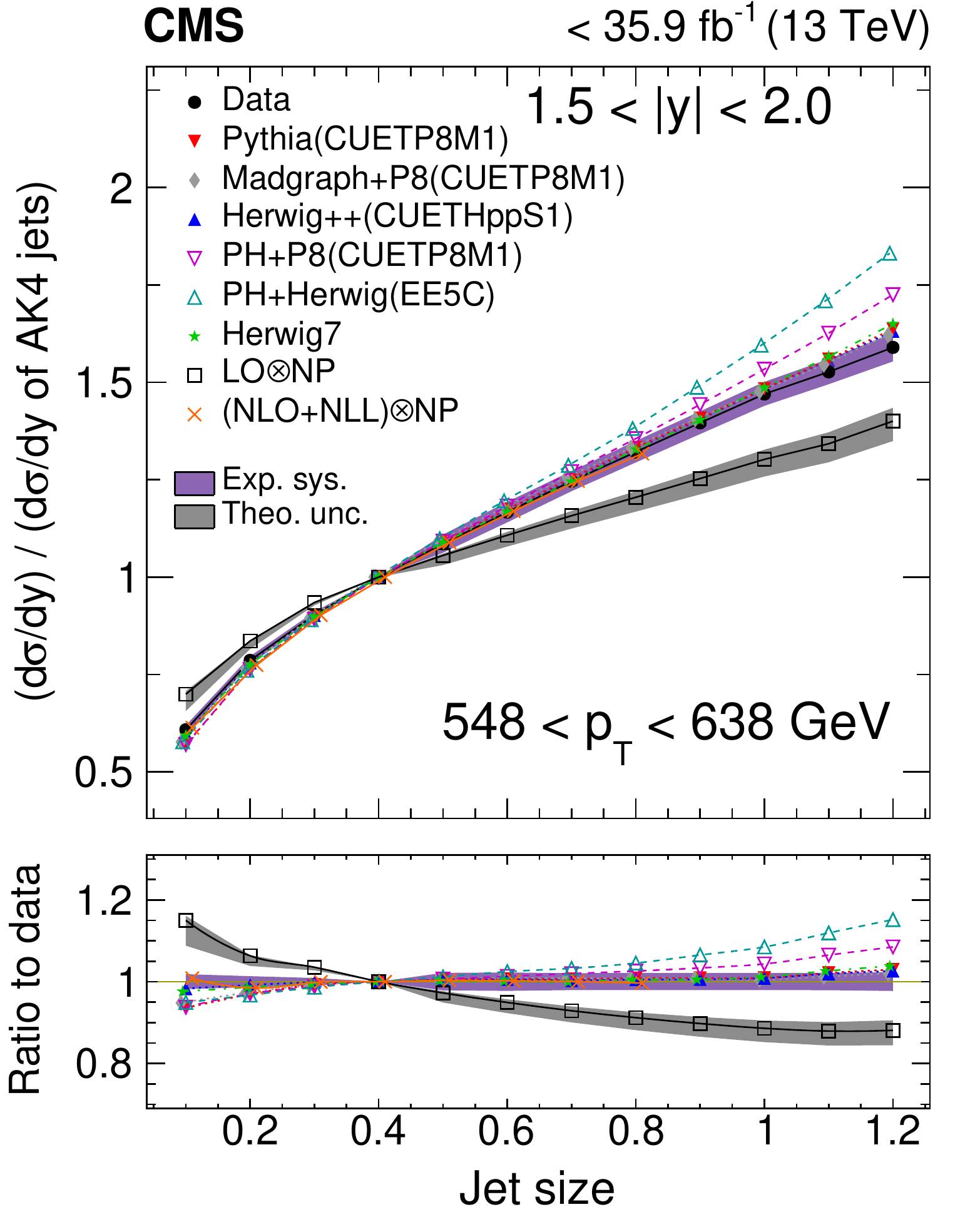}
\caption{Comparison of the ratio of cross sections of inclusive jets of various sizes with respect to AK4 jets, as a function of jet size in different regions of jet \pt in data, and for multiple theoretical predictions in rapidity bins $\abs{y}<0.5$ (left column) and $1.5<\abs{y}<2.0$ (right column) at particle level. 
When the dijet production cross section ratio is presented using pure NLO predictions for two jet sizes, the ratio becomes LO at \alpS ; this is quoted as LO$\otimes$NP in the figure. 
Points corresponding to a particular prediction are connected via lines to guide the eye.
Experimental uncertainties in the ratio of cross sections are shown with bands around the data points, whereas theoretical uncertainties are shown with the bands around the fixed-order predictions.
}
\label{fig:genptintfinal}
\end{figure}

The cross section is determined as a function of \pt for both data and theoretical predictions. 
The numbers are then divided by the cross section for the AK4 jets in the same \pt and rapidity window separately for data and each theoretical prediction, and  presented in Fig.~\ref{fig:genptintfinal}, in three ranges of \pt for the most central ($\abs{y}<0.5$) and the most forward ($1.5<\abs{y}<2.0$) regions as a function of jet size. Almost all the MC simulations involving resummation via parton shower can describe the trend with jet size seen in data, whereas the LO calculation exhibits different behavior.
Prediction from NLO calculation, as shown in $\abs{y}<0.5$,  improves significantly the description of cross section ratio, as observed in data, for small jet sizes, and lies between the LO prediction and data for large jet sizes.
Analytic calculations with joint resummation, available for jet sizes up to 0.8, provide an advancement with respect to fixed-order predictions, and lead to a better agreement with data. 
Similar behavior is observed in all the rapidity regions reported. 

\section{Summary}

A measurement has been made of the ratio of cross sections of inclusive anti-\kt jets of multiple sizes with respect to jets with the distance parameter $R = 0.4$;  this is the first such result from the CMS Collaboration. Because of cancellation of many experimental and theoretical systematic uncertainties for the ratio, it is more sensitive to perturbative and nonperturbative effects than the absolute cross section measurement; the experimental systematic uncertainty in the cross section ratio is of similar size as the statistical uncertainty, whereas the theoretical uncertainty is dominated by the choice of the renormalization and factorization scales. 

From the ratio measurement, we observe that the nonperturbative correction is important in describing the data at low transverse momentum. 
Thus, the modeling of nonperturbative effects, such as hadronization and the underlying event has a significant impact on the description of the data in different regions of phase space.

Finally, the variation of the ratio of cross sections with jet size $R$ emphasizes the importance of the inclusion of parton showering algorithms to capture the effects of higher-order terms in the
perturbation series by the resummation approach, which are absent in the case of fixed-order computation.
This is also demonstrated by the analytic calculations using joint resummation in threshold for single jet production, and jet size.
 Therefore, this study shows the importance of final-state radiation modeled in Monte
Carlo simulation to describe the data, and also implies that the differences between various parton showering and hadronization models are significant.

\begin{acknowledgments}

  We congratulate our colleagues in the CERN accelerator departments for the excellent performance of the LHC and thank the technical and administrative staffs at CERN and at other CMS institutes for their contributions to the success of the CMS effort. In addition, we gratefully acknowledge the computing centers and personnel of the Worldwide LHC Computing Grid for delivering so effectively the computing infrastructure essential to our analyses. Finally, we acknowledge the enduring support for the construction and operation of the LHC and the CMS detector provided by the following funding agencies: BMBWF and FWF (Austria); FNRS and FWO (Belgium); CNPq, CAPES, FAPERJ, FAPERGS, and FAPESP (Brazil); MES (Bulgaria); CERN; CAS, MoST, and NSFC (China); COLCIENCIAS (Colombia); MSES and CSF (Croatia); RPF (Cyprus); SENESCYT (Ecuador); MoER, ERC IUT, PUT and ERDF (Estonia); Academy of Finland, MEC, and HIP (Finland); CEA and CNRS/IN2P3 (France); BMBF, DFG, and HGF (Germany); GSRT (Greece); NKFIA (Hungary); DAE and DST (India); IPM (Iran); SFI (Ireland); INFN (Italy); MSIP and NRF (Republic of Korea); MES (Latvia); LAS (Lithuania); MOE and UM (Malaysia); BUAP, CINVESTAV, CONACYT, LNS, SEP, and UASLP-FAI (Mexico); MOS (Montenegro); MBIE (New Zealand); PAEC (Pakistan); MSHE and NSC (Poland); FCT (Portugal); JINR (Dubna); MON, RosAtom, RAS, RFBR, and NRC KI (Russia); MESTD (Serbia); SEIDI, CPAN, PCTI, and FEDER (Spain); MOSTR (Sri Lanka); Swiss Funding Agencies (Switzerland); MST (Taipei); ThEPCenter, IPST, STAR, and NSTDA (Thailand); TUBITAK and TAEK (Turkey); NASU (Ukraine); STFC (United Kingdom); DOE and NSF (USA). 
  
  \hyphenation{Rachada-pisek} Individuals have received support from the Marie-Curie program and the European Research Council and Horizon 2020 Grant, contract Nos.\ 675440, 752730, and 765710 (European Union); the Leventis Foundation; the A.P.\ Sloan Foundation; the Alexander von Humboldt Foundation; the Belgian Federal Science Policy Office; the Fonds pour la Formation \`a la Recherche dans l'Industrie et dans l'Agriculture (FRIA-Belgium); the Agentschap voor Innovatie door Wetenschap en Technologie (IWT-Belgium); the F.R.S.-FNRS and FWO (Belgium) under the ``Excellence of Science -- EOS" -- be.h project n.\ 30820817; the Beijing Municipal Science \& Technology Commission, No. Z191100007219010; the Ministry of Education, Youth and Sports (MEYS) of the Czech Republic; the Deutsche Forschungsgemeinschaft (DFG) under Germany's Excellence Strategy -- EXC 2121 ``Quantum Universe" -- 390833306; the Lend\"ulet (``Momentum") Program and the J\'anos Bolyai Research Scholarship of the Hungarian Academy of Sciences, the New National Excellence Program \'UNKP, the NKFIA research grants 123842, 123959, 124845, 124850, 125105, 128713, 128786, and 129058 (Hungary); the Council of Science and Industrial Research, India; the HOMING PLUS program of the Foundation for Polish Science, cofinanced from European Union, Regional Development Fund, the Mobility Plus program of the Ministry of Science and Higher Education, the National Science Center (Poland), contracts Harmonia 2014/14/M/ST2/00428, Opus 2014/13/B/ST2/02543, 2014/15/B/ST2/03998, and 2015/19/B/ST2/02861, Sonata-bis 2012/07/E/ST2/01406; the National Priorities Research Program by Qatar National Research Fund; the Ministry of Science and Education, grant no. 14.W03.31.0026 (Russia); the Tomsk Polytechnic University Competitiveness Enhancement Program and ``Nauka" Project FSWW-2020-0008 (Russia); the Programa Estatal de Fomento de la Investigaci{\'o}n Cient{\'i}fica y T{\'e}cnica de Excelencia Mar\'{\i}a de Maeztu, grant MDM-2015-0509 and the Programa Severo Ochoa del Principado de Asturias; the Thalis and Aristeia programs cofinanced by EU-ESF and the Greek NSRF; the Rachadapisek Sompot Fund for Postdoctoral Fellowship, Chulalongkorn University and the Chulalongkorn Academic into Its 2nd Century Project Advancement Project (Thailand); the Kavli Foundation; the Nvidia Corporation; the SuperMicro Corporation; the Welch Foundation, contract C-1845; and the Weston Havens Foundation (USA). 

\end{acknowledgments}

\bibliography{auto_generated}
\cleardoublepage \appendix\section{The CMS Collaboration \label{app:collab}}\begin{sloppypar}\hyphenpenalty=5000\widowpenalty=500\clubpenalty=5000\vskip\cmsinstskip
\textbf{Yerevan Physics Institute, Yerevan, Armenia}\\*[0pt]
A.M.~Sirunyan$^{\textrm{\dag}}$, A.~Tumasyan
\vskip\cmsinstskip
\textbf{Institut f\"{u}r Hochenergiephysik, Wien, Austria}\\*[0pt]
W.~Adam, F.~Ambrogi, T.~Bergauer, J.~Brandstetter, M.~Dragicevic, J.~Er\"{o}, A.~Escalante~Del~Valle, M.~Flechl, R.~Fr\"{u}hwirth\cmsAuthorMark{1}, M.~Jeitler\cmsAuthorMark{1}, N.~Krammer, I.~Kr\"{a}tschmer, D.~Liko, T.~Madlener, I.~Mikulec, N.~Rad, J.~Schieck\cmsAuthorMark{1}, R.~Sch\"{o}fbeck, M.~Spanring, D.~Spitzbart, W.~Waltenberger, C.-E.~Wulz\cmsAuthorMark{1}, M.~Zarucki
% ...
\vskip\cmsinstskip
\dag: Deceased\\
1:  Also at Vienna University of Technology, Vienna, Austria\\
% ...
\end{sloppypar}
\end{document}